\documentclass[12pt]{iopart}

\usepackage{iopams}  
\usepackage{graphicx}
\usepackage{float}
\usepackage{dcolumn}
\usepackage{bm}
\usepackage{cite}
\usepackage{color}
\begin{document}

\title{Pattern-induced anchoring transitions in nematic liquid crystals}

\author{\'Oscar A. Rojas-G\'omez}
\address{Departamento de F\'{\i}sica At\'omica, Molecular y
Nuclear, Area de F\'{\i}sica Te\'orica, Universidad de Sevilla,
Apartado de Correos 1065, 41080 Sevilla, Spain}
\author{Jos\'e M. Romero-Enrique}
\address{Departamento de F\'{\i}sica At\'omica, Molecular y
Nuclear, Area de F\'{\i}sica Te\'orica, Universidad de Sevilla,
Apartado de Correos 1065, 41080 Sevilla, Spain}
\author{Nuno M. Silvestre}
\address{Departamento de F\'{\i}sica e Centro de F\'{\i}sica Te\'orica e Computacional, 
Faculdade de Ci\^encias, Universidade de Lisboa, Campo Grande, P-1749-016, Lisbon, Portugal}
\author{Margarida M. Telo da Gama}
\address{Departamento de F\'{\i}sica e Centro de F\'{\i}sica Te\'orica e Computacional, 
Faculdade de Ci\^encias, Universidade de Lisboa, Campo Grande, P-1749-016, Lisbon, Portugal}
\begin{abstract}
In this paper we revisit the problem of a nematic liquid crystal in contact 
with patterned substrates. The substrate is modelled as a periodic array of 
parallel infinite grooves of well-defined cross section sculpted on a 
chemically homogeneous substrate which favors local homeotropic anchoring 
of the nematic. We consider three cases: a sawtooth, a crenellated and a sinusoidal substrate. 
We analyse this problem within the 
modified Frank-Oseen formalism. We argue that, for substrate periodicities much larger 
than the extrapolation length, the existence of different nematic textures 
with distinct far-field orientations, as well as the anchoring transitions 
between them, are associated with the presence of topological defects either on 
or close to the substrate. For the sawtooth and sinusoidal case, we observe a 
homeotropic to planar anchoring transition as the substrate roughness is increased. On the other hand, 
a homeotropic to oblique anchoring transition is observed for crenellated substrates. In this case,
the anchoring phase diagram shows a complex dependence on the substrate roughness and substrate anchoring strength.
\end{abstract}

\maketitle

\section{Introduction\label{sec1}}

In the last decades the study of nematic liquid crystals in the presence of microstructured
substrates has been the subject of intense research \cite{lee,kim,ferjani}.
This problem is interesting not only from a fundamental point of view, but also due to its
practical applications, such as the design of
zenithally bistable devices \cite{Brown_2000,Parry-Jones1,Parry-Jones2,
davidson1,evans,dammone}, or the trapping of colloidal particles on specified sites
\cite{nuno,ohzono,nuno2,nuno3,luo}. The presence of the structured substrate typically distorts the nematic 
orientational order, leading to elastic distortions and the formation of topological defects. 
On the other hand, the substrate topography can determine the director orientation far away from the substrate.
Since the seminal work of Berreman
\cite{berreman,degennes}, this problem has been extensively studied and
generalized in the literature \cite{barbero1,barbero2,barbero3,Brown_2000,
Kitson_2002,Fukuda,patricio_dietrich,kondrat,harnau,kondrat2,harnau2,harnau3,barbero,Yi_2009,
poniewierski,patricio2,oscar,raisch,ledney}. Wetting and filling transitions by nematics on these
surfaces have also been studied \cite{bramble,patricio1,patricio3,
patricio4,patricio0,patricio5}. 
When the substrate has cusps, disclination-like singularities generally
appear at or very close to them \cite{barbero1,barbero2,barbero3,poniewierski,patricio2,oscar,lewis}. However, even when 
the substrate is smooth disclination lines may appear, in the nematic, close to the substrate \cite{patricio4,raisch,ledney}.
In both cases, these orientation-field singularities play an important role to understand the different 
textures of the nematic in contact with a patterned substrate under strong anchoring conditions.

In this paper we investigate the equilibrium nematic texture at substrates of arbitrary cross section,
and determine how this texture determines the (homogeneous) nematic director in the far-field. Anchoring transitions
are identified as the transitions between nematic textures with different far-field orientations. 
We will assume that the nematic director field lies in the plane perpendicular to the patterned substrate longitudinal axis, 
and that the substrate lengthscales are large enough to ensure strong anchoring conditions on the substrate.   
In a few cases analytical results are available \cite{barbero1,Yi_2009,oscar,lewis,ledney,davidson}, but in general we have
to resort to numerical methods. 

Most studies involve numerical minimization of free-energy functionals such as 
the Landau-de Gennes or Frank-Oseen models. The Landau-de Gennes model describes the emergence of topological defects, but
it is very time-consuming and it is difficult to obtain systematic results when the pattern length scales are much larger than the 
nematic coherence length. On the other hand, in the Frank-Oseen model topological defects need to be included by
hand. 

In Ref. \cite{oscar} we extended the Frank-Oseen model to include disclination-like singularities near the 
cusps of a sawtooth substrate, and found excellent agreement with previous Landau-de Gennes calculations \cite{patricio2}. 
In this paper we generalize this numerical method to surface reliefs of arbitrary sections
(with or without cusps), as well as in the presence of disclination lines in the nematic phase. We apply our method
to study the anchoring transitions induced by the patterned substrate in sawtooth, crenellated and sinusoidal substrates
under strong anchoring conditions. 
    
The paper is organized as follows. In section \ref{sec2} is devoted we set the problem. The numerical technique is 
presented in Section \ref{sec3}. Results on sawtooth, crenellated and sinusoidal substrates are discussed in
Section \ref{sec4}. Finally, we end with the conclusions in Section \ref{sec5}.
Technical details of the numerical method and of the analytic 
solution of the free-energy elastic contribution for a particular nematic 
texture on crenellated substrates are described in the appendix.

\section{The model \label{sec2}}
We consider a nematic liquid in contact with a patterned substrate with a relief 
profile $\psi(x,z)$ that favours local homeotropic anchoring of the molecules
(see Fig. \ref{fig1}). Translational symmetry along the $z$ axis is assumed, so that
$\psi=\psi(x)$. On the other hand, the substrate is periodic along the $x$-axis with 
a wavelength $\lambda$, i.e. $\psi(x+\lambda)=\psi(x)$. 
Furthermore, we assume that the nematic director field $\mathbf{n}(\mathbf{r})$ exhibits only 
in-plane distortions, and thus it can be parametrized by the angle $\theta$ between
the local director and the $y$ axis, yielding
$\mathbf{n}(\mathbf{r})=(-\sin\theta(\mathbf{r}),\cos\theta(\mathbf{r}),0)$.
The nematic order may be represented locally by a traceless
symmetric second-rank tensor order parameter $\mathbf{Q}$, with Cartesian components
$Q_{ij} = \frac 3 2 S[n_in_j - \frac 1 3 
\delta_{ij}]+ \frac 1 2 B [l_il_j - m_i m_j]$, where $S$ is the nematic
order parameter, which measures the orientational ordering along the
nematic director, and $B$ the biaxiality order parameter, which
measures the ordering of the molecules along directions 
perpendicular to $\mathbf{n}$, characterized by the eigenvectors
$\mathbf{l}$ and $\mathbf{m}$. We consider uniaxial nematic liquid crystals, so that $B=0$ 
except close to the substrate or within the topological defect cores. 
Note that an inversion of $\mathbf{n}$ does not change the value of $\mathbf{Q}$, and thus
in nematics, $\mathbf{n}$ and $-\mathbf{n}$ are physically equivalent configurations.

Far from the substrate, no specific orientation is imposed, but we require that the 
bulk nematic phase is oriented uniformly along some direction not specified, implying that
$\boldsymbol{\nabla}\theta \to \mathbf{0}$ as $y\to \infty$. We anticipate that 
different nematic textures close to the patterned surface will lead, in general, to 
distinct nematic far-field orientations, and thus surface transitions will change the 
nematic anchoring with respect to the substrate
reference plane $xz$. In general different textures have different symmetries, and thus the
anchoring transitions are expected be first-order. Close to the transitions the surface states are locally
stable in a thermodynamic sense, and these geometries are ideal candidates for the design of zenithally bistable devices. 

The excess free energy $F$ can be written as $F_e+F_a+F_c$,  where $F_e$ is the elastic 
contribution, $F_a$ is the free-energy contribution associated to the anchoring of the
nematogen molecules to the substrate and $F_c$ is the contribution
associated to the disclination cores. $F_e$ is given by the Frank-Oseen elastic free 
energy \cite{oseen,frank}:
\begin{eqnarray}
F_e &=& \frac{1}{2}\int_{\cal V} d\mathbf{r} \Bigg[K_1 (\boldsymbol{\nabla\cdot}\mathbf{n})^2
+K_2(\mathbf{n\cdot}\boldsymbol{\nabla}\times \mathbf{n})^2
\label{frank} \\&+&
K_3 (\mathbf{n}\times\boldsymbol{\nabla}\times\mathbf{n})^2
+ K_{24}
\boldsymbol{\nabla\cdot}
[(\mathbf{n\cdot} \boldsymbol{\nabla})\mathbf{n}-\mathbf{n}
(\boldsymbol{\nabla\cdot}\mathbf{n})]\Bigg]
\nonumber
\end{eqnarray}
where ${\cal V}$ is the volume occupied by the nematic liquid crystal, 
$K_1$, $K_2$ and $K_3$ are the splay, twist and bend bulk elastic 
constants, respectively, and $K_{24}$ is the saddle-splay elastic constant.
Thus, using
the parametrization of $\mathbf{n}$ in terms of the field $\theta(x,y)$, the elastic
contribution to the nematic free energy per unit length along the $z$ axis, $f_e=F_e/L_z$, 
is:
\begin{equation}
f_e=\frac{K_1}{2}\int_{\cal A} 
\left(|\boldsymbol{\nabla}\theta|^2+ \left(\frac{K_3-K_1}{K_1}\right)(\mathbf{n\cdot}
\boldsymbol{\nabla}\theta)^2 \right)dxdy
\label{frank2}
\end{equation}
where ${\cal A}$ is the $xy$ section of the volume occupied by the nematic liquid crystal. 
Note that the twist and saddle-splay contributions vanish
identically. If we further assume that the splay and bend elastic constants are equal,
i.e. $K_1=K_3=K$, then:
\begin{equation}
f_e=\frac{K}{2}\int_{\cal A} 
|\boldsymbol{\nabla}\theta|^2 dxdy
\label{frank3}
\end{equation} 
The anchoring free-energy contribution $F_a$ is modelled by the Rapini-Papoular 
approximation \cite{rapini}
\begin{equation}
F_a=-\frac{W}{2}\int_{\cal S} d\mathbf{s} (\mathbf{n}(\mathbf{s})\mathbf{\cdot}
\boldsymbol{\nu}(\mathbf{s}))^2
\label{rapini}
\end{equation}
where the integral is over the substrate surface ${\cal S}$, and 
$\boldsymbol{\nu}(\mathbf{s})$ is the outwards unit vector normal to the substrate at 
$\mathbf{s}$. 
Thus, the anchoring contribution to the nematic free energy per unit length along the $z$
axis, $f_a=F_a/L_z$, is:
\begin{equation}
f_a=-\frac{W}{2}\int_{\cal L} ds \cos^2 \psi(s)
\label{rapini2}
\end{equation}
where ${\cal L}$ is the curve $(x,\psi(x),0)$, parametrized by its natural parameter $s$,
and $\psi$ is the angle between $\mathbf{n}$ and $\boldsymbol{\nu}$, i.e. $\psi=\theta-\alpha$,
with $\alpha(x)=\arctan(\psi'(x))$.

In order to obtain the equilibrium nematic texture, we have to minimize the functional
$f\equiv f[\theta]=f_e+f_a+f_c$, with $f_e$ and $f_a$ given by Eqs. (\ref{frank3}) and 
(\ref{rapini2}), respectively. Due to the symmetries of the problem, we have to find the 
solution only in the $xy$ region $R$ delimited by the dashed line in Fig. \ref{fig2} and 
bounded below by one period of the substrate relief. The Euler-Lagrange equation 
associated to the functional $f[\theta]$ reduces to the Laplace equation in $R$, 
$\nabla^2 \theta=0$, subject to the boundary condition on the substrate:
\begin{equation}
\xi_{ext}\boldsymbol{\nu\cdot\nabla}\theta(s)=-\frac{1}{2}\sin2\psi(s) \equiv -\frac{1}{2}
\sin 2(\theta(s)-\alpha(s))
\label{bc1}
\end{equation}
where $\xi_{ext}\equiv K/W$ is the Kl\'eman-de Gennes extrapolation length \cite{degennes}.  
In addition, periodic boundary conditions are imposed on the sides $x=-\lambda/2$ and 
$x=\lambda/2$, and free (Neumann) boundary conditions $\partial \theta/\partial y (y=H)=0$ at the far field. 

Some comments are in order at this point. First, other boundary
conditions on the lateral sides are allowed. For example, we can assume that $\theta(x+\lambda,y)=\theta(x,y)+m\pi$,
where $m$ is an integer. However, this condition leads to a solution of the Laplace equation
which is linear in $x$ far away from the substrate, leading to an infinite interfacial free energy as a result of
elastic distortions in the bulk. As we are interested in the anchoring of an undistorted bulk nematic at a patterned 
substrate, we discard these solutions. On the other hand, we take 
the limit $H\to \infty$. Finite values of $H$ are relevant to study nematic textures under confinement.

The continuum model described above is valid as long as all the characteristic relief
lengths are much larger than the nematic coherence length $\xi_0$, which is of the order 
of the molecular size. For strong anchoring conditions, $\xi_{ext}\sim \xi_0$ and 
Eq. (\ref{bc1}) can be substituted by the strong anchoring condition $\theta(s)=\alpha(s)$
on the substrate. For weak anchoring, $\xi_{ext}\gg \xi_0$, and in principle this lengthscale
is also relevant. In typical liquid crystals $\xi_0\sim 1$ nm and 
$\xi_{ext}\sim 0.1-10\ \mu m$. However, if we assume that the typical lengthscale 
which controls the substrate pattern is $\lambda$, then strong anchoring conditions on the
substrate may be considered also for $\xi_{ext}\gg \xi_0$ if 
$\lambda\gg \xi_{ext}$. We can justify this result 
by rescaling the domain by a factor $\lambda$ \cite{patricio2}: 
$\mathbf{r}^*=\mathbf{r}/\lambda$ and $\theta^*(\mathbf{r}^*)=\theta(\mathbf{r})$.
In order to minimize the free energy functional in this rescaled description, we  
solve the Laplace equation in the rescaled domain subject to an effective
anchoring potential $W^*=\lambda W$ on the substrate, so the rescaled extrapolation
length is $\xi_{ext}^*=\xi_{ext}/\lambda\ll 1$ in this limit. 

The question about the existence of anchoring transitions in the strong anchoring limit is
related to the uniqueness of the solutions of the Laplace equation subject to the boundary 
conditions described above. A standard analysis of the mathematical problem shows that
the Laplace equation is unique for a given integration domain and boundary conditions on the
substrate. In fact, the eigenvalues of the Laplacian with the corresponding  
homogeneous boundary conditions are strictly positive, as inferred from its
Rayleigh quotient, and the uniqueness of the solution is guaranteed via Fredholm's alternative. 
Thus, only one texture is expected under strong anchoring conditions. 
This result is consistent with previous results in the literature 
\cite{kondrat,harnau,barbero}, where it
is shown that for patterned and/or chemically heterogeneous substrates a unique nematic
texture is observed if the extrapolation lengths are much larger than $\lambda$. 
On the other hand, a transition between an almost homogeneous and a distorted texture
may be observed if any of the relevant extrapolation lengths is of order of $\lambda$
\cite{kondrat,harnau}. The driving force for this transition is the competition between 
the elastic and the anchoring contributions to the free energy. This mechanism
is not relevant at large $\lambda$, since $f_e \sim K$ while $f_a \sim \lambda W$ 
when $\lambda$ is large. 

However, there is  analytical, numerical and experimental evidence that different textures 
are indeed possible for a given substrate relief, even in the strong anchoring limit
\cite{davidson1,luo,barbero1,barbero2,barbero3,patricio2,poniewierski,oscar,lewis,ledney}. The 
apparent contradiction with the previous result may be resolved by noting that 
some of the textures exhibit topological defects in the nematic, i.e. $\pm 1/2$ 
disclination lines or, in their absence, the relief has cusps. 
In the first situation, the presence of a disclination line in the nematic modifies the
domain where the Laplace equation is solved, since the solution has a singularity at the
defect core. In addition, a branch cut must be added in order to avoid the non-physical
(but mathematically sound) jump of $\theta$ by $\pm \pi$ when following a loop  
enclosing the defect core. Thus the domain where the Laplace equation is solved has to be 
modified by excluding both the defect core and the branch cut, with additional conditions 
on these new boundaries. Furthermore, the free energy per unit length
will have a contribution $f_c$ due to the destruction of orientational order at the core of the 
disclination line. 

On the other hand, when the substrate relief has cusps, 
the angle $\theta$ exhibits a discontinuity under strong anchoring conditions. This implies that 
the solution has a disclination-like singularity close to the cusp, with
its associated core contribution $f_c$ \cite{oscar}. The strength of the singularity $I$, i.e. its 
effective topological charge, is related geometrically to
the opening angle of the cusp $\Delta \phi$ and the jump $\Delta \theta$ of the orientational
field $\theta$ when crossing the cusp along the surface from right to left as $I=\Delta 
\theta/\Delta \phi$ \cite{patricio5}. One posibility is that $\Delta \theta_0=\Delta \phi-
\pi$, which leads to $I_0=1-\pi/\Delta \phi$. However, this is not the only option
due to the physical 
equivalence between $\mathbf{n}$ and $-\mathbf{n}$. In particular, it is posible
that $\Delta \theta_m=\Delta \theta_0 + m\pi$ for $m\in \mathbb{Z}$, which leads to $I_m=1+(m-1)\pi/\Delta\phi$.
In each case, the boundary conditions are
mathematically different, although physically equivalent. Thus different textures may occur for the same substrate 
under strong anchoring conditions. 
 
The elastic energy per unit length along the $z$ axis 
has contributions associated to the disclination lines and the distortions close to the cusps of the form
\cite{patricio5}:
\begin{equation}
\frac{K}{2}\left(\frac{N_D\pi}{2}+\sum_{i}I_i^2 \Delta \phi_i\right)  \ln \left(\frac{\lambda}{\xi_0}\right)
\label{asymptotic-core}
\end{equation}
where $N_D$ is the total number of disclination lines in the nematic and 
the index $i$ runs over all the cusps on the substrate. The next-to-leading
contribution to the free energy per unit length is expected to be independent of the lengthscale $\lambda$. 
Thus, for large $\lambda$ we expect only
transitions between textures with the same leading-order contribution to the free energy given by Eq. 
(\ref{asymptotic-core}). However, we will see that for moderate values of $\lambda$ other transitions between
textures may be observed.

\section{Numerical method \label{sec3}}

Analytical solutions of nematic textures in contact with patterned 
substrates are not available, in general. There are exceptions, which may be obtained, for example, by using conformal mapping techniques \cite{davidson}. 
Otherwise,
we have to resort to numerical methods. However, the presence of disclination lines and/or
singularities associated to the surface cusps requires special techniques.
In this section we will describe the method used 
to obtain the nematic textures in the presence of disclination lines and/or substrate cusps,
which is based on the numerical methods used previously for the sawtooth substrate
\cite{oscar}. First, we will assume that the number and positions of the disclination
lines in the nematic are known. In this case, the elastic contribution will be a function of the 
number of disclination lines and their corresponding positions. Their equilibrium values can be obtained by
standard minimization techniques, such as conjugated-gradient methods. 

We split the orientation field $\theta(x,y)$ in 
two terms: a singular contribution $\theta_s$, associated to the disclination lines in 
the nematic and/or the 
disclination-like singular contributions due to the surface cusps, and $\theta_{ns}$ which
we require to be regular everywhere in the integration domain $R$. If disclination lines 
are present in the nematic, we modify the integration $R$ to $\overline{R}$, excluding
the defect core and the branch cut, which we will consider perpendicular as shown in Fig. 
\ref{fig3}. We choose $\theta_s$ periodic in $x$, with period $\lambda$,  
satisfying the Laplace equation in $\overline{R}$ ($R$ in the absence of disclination lines),
which captures the singularities in the orientational field associated to the disclination lines 
and surface cusps. A possible choice, based in previous studies for
the sawtooth substrate, is \cite{patricio2,oscar}:
\begin{eqnarray}
\theta_s&=&\sum_{i} I_i\Bigg[\arctan\left(\frac{\tanh\frac{q}{2}(y-\psi(x_i))}{\tan\frac{q}{2}(x-x_i)}\right)\label{thetas}\\
&-&\arctan\left(\frac{1}{\tan\frac{q}{2}(x-x_i)}\right)\Bigg]
\nonumber\\
&+& \sum_j I_j \Bigg[\arctan\left(\frac{\tanh\frac{q}{2}(y-y_j)}{\tan\frac{q}{2}(x-x_j)}\right)\nonumber\\
&+&\arctan\left(\frac{1}{\tan\frac{q}{2}(x-x_j)}\right)\Bigg] 
\nonumber
\end{eqnarray}
where $q=2\pi/\lambda$, the first sum runs over the surface cusps at $(x_i,\psi(x_i))$, while the second one is over the 
disclination lines at positions $(x_j,y_j)$ in the nematic phase. Note that the contribution associated to the surface 
cusps vanishes as $y\to \infty$. By contrast, the nematic disclination term exhibits a piecewise linear behaviour
as $y\to \infty$ and its elastic contribution to the interfacial free energy diverges unless the total topological charge 
associated to the nematic disclination lines vanishes. Therefore, we restrict our study to situations where the number of $+1/2$-disclination lines in the nematic phase is the
same as the number of $-1/2$-disclination lines to ensure that the far-field director field is undistorted.

For the non-singular part $\theta_{ns}$, 
we have to solve the Laplace equation subject to periodic boundary conditions on the sides
$x=\pm\lambda/2$ and free boundary conditions as $y\to \infty$. Finally, $\theta_{ns}$ satisfies Dirichlet boundary 
conditions $\theta_{ns}(s)=\alpha(s)-\theta_s (s)$ at the substrate . As mentioned above, the regularity
of $\theta_{ns}$ in the region $R$ implies that standard numerical techniques can be used. We use
the boundary element method to obtain $\theta_{ns}$ in the constant element approximation \cite{brebbia,katsikadelis}.
In Ref. \cite{oscar} we
used a boundary-element method where the whole boundary of $R$ is discretized. In this paper we use a different 
boundary-element method that requires only the discretization of the substrate relief. A detailed 
description of this technique can be found in \ref{appendixa}. For this purpose, a polygonal approximation to the 
substrate relief is considered, where each segment length is small with respect to $\lambda$ (the only lengthscale 
relevant for this problem). Thus, for the sawtooth and sinusoidal substrates, the substrate is divided into 720 segments with
the same $x$-axis projection length. On the other hand, for the crenellated substrate each side of the substrate 
is divided into 120 segments of the same length. We checked that our numerical results are, within numerical accuracy,
almost identical when finer discretizations are considered. 

Once the orientational field is obtained, we need to evaluate the corresponding interfacial free energy.
The elastic contribution to the free energy per $z-$unit length and $x-$period $f_e$ 
can be obtained from Eq. (\ref{frank3}) as:
\begin{equation}
f_e=\frac{K}{2}\int_{\cal A} |\boldsymbol{\nabla} \theta|^2 d\mathbf{r}=
\frac{K}{2} \oint \theta (\boldsymbol{\nu\cdot\nabla}\theta) d\mathbf{s}\label{energy}
\label{contourfreeenergy}
\end{equation}
Technical details on how to evaluate this contribution from the singular and non-singular parts of the
orientational field are described in \ref{appendixb}.

The final ingredients are the core contributions
$f_c$ arising from the disclination lines and the effective 
disclination-like singularities of the nematic 
textures. These terms are not described by the macroscopic elastic theory, and we have to resort to a more microscopic 
description to
evaluate them. We will evaluate them using the mesoscopic Landau-de Gennes 
framework $F_{LdG}=\int_{\cal V}{d\mathbf{r}
\left({\cal F}_b\left(\mathbf{Q}\right)+
{\cal F}_e\left(\partial\mathbf{Q}\right)\right)}+\int_{\cal S} {\cal F}_s(\mathbf{Q})$, 
where the bulk and elastic free energy densities are, respectively,
\begin{eqnarray}
 {\cal F}_b&=&a_o\left(T-T^*\right)\Tr \mathbf{Q}^2-b\Tr\mathbf{Q}^3+
c\left(\Tr\mathbf{Q}^2\right)^2\\
 {\cal F}_e&=&\frac{L_1}{2}\partial_\gamma Q_{\alpha\beta}\partial_\gamma Q_{\beta\alpha}
 + \frac{L_2}{2}\partial_\gamma Q_{\alpha\gamma}\partial_\delta Q_{\delta\alpha}.
 \end{eqnarray}
The bulk term ${\cal F}_b$ determines the bulk nematic order parameter: $S=0$ 
(isotropic phase) if $\tau=24a_o(T-T^*)c/b^2>1$, and 
$S=\left(b/8c\right)\left(1 + \sqrt{1-8\tau/9}\right)$ (nematic phase)
if $\tau<1$. The elastic term ${\cal F}_e$ penalizes distorsions of the 
orientational field, with two elastic constants $L_1$ and $L_2$ 
related to the Frank-Oseen elastic constants: $K_{1}=K_{3}=9S^2L_1
\left(2+L_2/L_1\right)/4$ and $K_2=9S^2L_1/2$.
In addition, we consider the surface free energy density used in Refs. 
\cite{patricio1,patricio2,patricio3,patricio4,patricio5,oscar}:
\begin{equation}
{\cal F}_s = -w \Tr \mathbf{Q\cdot Q}_s
\end{equation}
where $w$ is a parameter related to the anchoring strength \cite{oscar}
and $\mathbf{Q}_s$ is the reference tensor order parameter on the substrate with 
Cartesian components $(Q_s)_{ij}=(3\nu_i \nu_j - \delta_{ij})/2$, and $\nu_i$
the Cartesian components of the unit vector normal to the substrate
$\boldsymbol{\nu}$. We obtain the core contributions by using an adaptive-meshing finite-element method
combined with a conjugate-gradient minimization algorithm, following the procedure 
described in Ref. \cite{oscar}. Note that, as the typical size of the cores is
$\xi_0$, the cores associated with the disclination lines in the nematic are independent 
from the substrate. This is not the case at the cusps where we assume that two locally planar
surfaces meet. The core contribution of the cusp singularities 
depends, in general, on the anchoring strength.
In what follows, we take the nematic to be at nematic-isotropic 
coexistence (i.e. $\tau=1$). 
\section{Results \label{sec4}}

In this Section we will describe the results for different substrate reliefs. In particular, we will revisit 
the sawtooth case, and we will present results for the crenellated and the sinusoidal substrates.

\subsection{The sawtooth substrate}
The sawtooth substrate has been studied for a number of years \cite{barbero,patricio2,oscar}. We revisit some
of the results reported previously. We consider a symmetric sawtooth, characterized by a tilt angle
$\alpha$ and a side length $L$, as shown in Fig. \ref{fig4}(a), so that $\lambda=2L\cos\alpha$. No bulk disclinations 
are expected in the nematic texture, at least at distances of order $L$ from the substrate. 
It is found that there are
two nematic textures that are locally stable: the $N^\perp$ texture, where the nematic field is oriented
along the $y$ axis away from the substrate, in the far field, and the $N^\parallel$ texture, where the nematic field is 
oriented along the $x$ axis in the far field, as shown in Fig. \ref{fig5}. These textures are characterized by different effective topological
charges $I^t$ and $I^b$ associated to the top and bottom cusps, respectively. In the $N^\perp$ texture, 
$I^t=I^t_0\equiv\alpha/(\pi/2+\alpha)$ and $I^b=I^b_0\equiv-\alpha/(\pi/2-\alpha)$, while in the $N^\parallel$ texture, 
$I^t=I^t_{+1}\equiv-(\pi/2-\alpha)/(\pi/2+\alpha)$ 
and $I^b=I^b_{+1}\equiv 1$. The elastic contribution to the
interfacial free energy $f_e$ of the nematic (per $z-$ unit length and $x-$period) at this substrate was  
obtained analytically \cite{oscar}:
\begin{eqnarray}
f_e=\mathcal{K}_m(\alpha)\Bigg[-\ln \frac{q\xi_0\cos\alpha}{\pi}
- \left(\frac{1}{2}-\frac{\alpha}{\pi}\right)\ln\left(\frac{\frac{\pi}{2}+\alpha}
{\frac{\pi}{2}-\alpha}\right)\label{theoreticalf}\\
- \ln\left(\Gamma\left[\frac{3}{2}-\frac{\alpha}{\pi}
\right]\Gamma\left[\frac{1}{2}+\frac{\alpha}{\pi}\right]\right)\Bigg]
\nonumber
\end{eqnarray}
where $m=0$ for the $N^\perp$ texture and $m=+1$ for the $N^\parallel$ texture. The meaning of these
numbers will be discussed below. The effective elastic constants $\mathcal{K}_0$ and $\mathcal{K}_{+1}$ are
\begin{eqnarray}
\mathcal{K}_0(\alpha)=
\frac{K\pi \alpha^2}{\left(\frac{\pi}{2}
\right)^2-\alpha^2} \quad ; \quad  
\mathcal{K}_{+1}(\alpha)=K\pi\frac{\frac{\pi}{2}-\alpha}{\frac{\pi}{2}
+\alpha} 
\label{defkalpha}
\end{eqnarray}
Numerical results are in excellent agreement with the analytical result (see Ref.\cite{oscar} for a more detailed discussion).
As ${\mathcal K}_0(\pi/4)={\mathcal K}_{+1}(\pi/4)$, there is an anchoring transition from homeotropic to planar
anchoring at $\alpha=\pi/4$. However, the value of $\alpha$ at the transition may be altered in two ways. First, the core contributions
$f_c$ associated to the effective disclination-like singularities may be different for the $N^\perp$ and $N^\parallel$ 
textures. 
These core contributions will shift the transition (from $\alpha=\pi/4$) by a small amount, since $f_c\ll f_e \sim K\ln\lambda/\xi_0$ for large $\lambda$. Alternatively, for $\alpha=\pi/4$ this contribution drives the transition
between the $N^\perp$ and $N^\parallel$ when varying the value of the anchoring strength $w$.

The elastic constants anisotropy can also cause a shift in the anchoring transition. 
In the previous discussion we assumed that the splay and bend elastic constants are equal, in line with the observation that $K_3/K_1-1$ is very small for liquid crystals like 5CB, close to the nematic-isotropic phase transition. In order to estimate the effect of the elastic anisotropy on 
the anchoring 
transition, a perturbation theory around the one-elastic constant model was developed \cite{barbero}. The first
order correction in $K_3/K_1-1$ is obtained using Eq. (\ref{frank2}) \cite{barbero}:
\begin{equation} 
\frac{K_3-K_1}{2} \int_{\cal A} dxdy \left(\mathbf{n}_0\mathbf{\cdot}\boldsymbol{\nabla}\theta_0\right)^2
\label{perturbacion}
\end{equation}
where $\theta_0$ is the nematic orientation field for $K_1=K_3$ and $\mathbf{n}_0=(-\sin\theta_0,\cos\theta_0,0)$.
The main contribution arises from the neighbourhood of the cusps, which leads to an additional term proportional to 
$(K_3-K_1)\ln \lambda/\xi_0$. We refrain from giving explicit expressions for this leading-order correction, which implies that the anchoring transition shifts to $\alpha>\pi/4$ if $K_3>K_1$, and below $\pi/4$ otherwise. Physically this is due to the fact that the elastic distortions in the $N^\perp$ texture are mainly splay, while bend dominates in the $N^\parallel$ texture
(see Fig. \ref{fig5}). 

Finally, $N^\perp$ and $N^\parallel$ are not the only textures that are possible at a sawtooth substrate.
As mentioned in Section \ref{sec2}, the effective 
topological charges associated with the disclination-like singularities arising from
the substrate cusps may be expressed as $I_0+m\pi$, where $I_0$ is one possible value of the topological charge and 
$m$ is an integer. Therefore, there is an infinite number of (pairs) of topological charges $I^b$ and $I^t$, since $I^b_{m}=-\alpha/(\pi/2-\alpha)+m\pi/
(\pi-2\alpha)$ and $I^t_{m'}=\alpha/(\pi/2+\alpha)+m'\pi/(\pi-2\alpha)$ and the periodicity requirement on $\theta$ 
imposes 
$m'=-m$. The boundary conditions on $\theta$ are then $\theta=-\alpha$ on the left-to-right uphill segments,
and $\theta=-\alpha+m\pi$ on the downhill segments. The far-field value $\alpha_\infty$ is the average of these,
$\alpha_\infty=m\pi/2$, which is the value of $\theta$ along the vertical lines emerging from the substrate
cusps. Note that if $m=0$ and $m=+1$ we obtain the $N^\perp$ and the $N^\parallel$ textures, 
respectively, in line with the notation used in Eqs. (\ref{theoreticalf}) and (\ref{defkalpha}). 
The free energy of these nematic textures may be solved using the Schwarz-Christoffel conformal mapping used for the $N^\perp$ and $N^\parallel$ cases, leading to an elastic contribution
to the interfacial free energy given by Eq. (\ref{theoreticalf}), with $\mathcal{K}_m$ defined as:
\begin{eqnarray}
\mathcal{K}_m(\alpha)=\frac{K}{2} \left[\left(I_m^b\right)^2(\pi-2\alpha)+ \left(I_{-m}^t\right)^2 (\pi+2\alpha)\right]
\label{defkalpha2}\\
=\frac{K\pi}{\left(\frac{\pi}{2}\right)^2-\alpha^2}\left(m\frac{\pi}{2}-\alpha\right)^2=
\mathcal{K}_0(\alpha) \left(\frac{m\pi}{2\alpha}-1\right)^2
\nonumber
\end{eqnarray}
Fig. \ref{fig6} illustrates $\mathcal{K}_m$ as a function of $\alpha$. The lowest curves, with $m=0$ and $m=+1$, correspond to the $N^\perp$ and $N^\parallel$ textures, respectively. The other curves describe higher elastic energy states, and may be discarded at equilibrium. A similar behaviour was found for isolated wedges \cite{davidson}. 

\subsection{The crenellated substrate}
We proceed with the crenellated substrate, characterized by infinite blocks of width and height $l_1$ and $h$, respectively
at a distance $l_2$, as shown in Fig. \ref{fig4}(b). The substrate relief period is $\lambda=l_1+l_2$. 
As in the sawtooth, the presence of cusps in the substrate relief
leads to the appearance of disclination-like singularities in the orientational field nearby. 
By geometric considerations, the topological charges associated with the upper cusps (i.e. with opening angles
$3\pi/2$), $I^t_1$ and $I^t_2$, will be either $+1/3$ or $-1/3$,
and the charges associated with the lower cusps (i.e. with opening angles $\pi/2$), $I^b_1$ and $I^b_2$, will be
either $+1$ or $-1$. As in the sawtooth, not every combination
is possible due to the periodicity constraint. In order to ensure periodicity there must be two positive topological
charges (with the other two negative). Other values of the topological charges are possible, but as in the sawtooth case, 
they lead to much higher elastic free energies, which are irrelevant at equilibrium. 
Thus, we find 4 independent nematic textures:
$N^\perp_1$, where $I^t_1=I^t_2=-1/3$ and $I^b_1=I^b_2=+1$; $N^\perp_2$, with $I^t_1=I^t_2=+1/3$ and $I^b_1=I^b_2=-1$;
$N^o_1$ with $I^t_1=-1/3$, $I^b_1=+1$, $I^t_2=+1/3$ and $I^b_2=-1$, and finally $N^o_2$ with $I^t_1=-1/3$, $I^b_1=-1$, 
$I^t_2=+1/3$ and $I^b_2=+1$. Nematic textures obtained by the numerical minimization described in the previous section
are shown in Fig. \ref{fig7}. Note that both $N^\perp_1$ and $N^\perp_2$ are symmetric with respect to a mirror 
inversion, while $N^o_1$ and $N^o_2$ are asymmetric. Thus, for the latter 
there are two other equivalent textures related by mirror symmetry. With respect to the bulk nematic 
anchoring, the symmetric textures are homeotropic, i.e. the nematic director is oriented along the $y$ axis far away
from the substrate. The asymmetric textures, however,  exhibit oblique nematic anchoring. 
The far-field tilt angle $\alpha_\infty$ of the $N^o_1$ texture depends on $h/l_2$ and $l_1/l_2$.
For a given value of $l_1/l_2$ it increases monotonically with $h/l_2$ from zero and reaches a plateau at large
$h$ above $h/l_2\gtrsim 1$. The asymptotic values
of $\alpha_\infty$ at large $h/l_2$ decrease as $l_1/l_2$ increases, being almost proportional to $l_1/(l_1+l_2)$ at 
large $l_1$.  
Thus, narrow blocks lead to values of $\alpha_\infty\approx \pi/2$, while narrow channels lead to nearly homeotropic anchoring. 
Our numerical data also indicates that $\alpha_\infty$ satisfies approximately 
$\alpha(h/l_2;l_1/l_2)\approx \alpha_\infty(\infty;l_1/l_2)\chi(h/l_2)$ (see the inset of Fig. \ref{fig8}).
The existence of a plateau in $\alpha_\infty$ at large $h$ can be explained by noting that the nematic director in 
the region between the blocks at height $y\lesssim h-l_2$ (provided that
$h\gg l_2$) is almost the same as that in a rectangular well \cite{davidson}. This solution 
becomes almost parallel to the $x$ axis for $l_2\lesssim y -l_2$. In this case, the dependence on $h$ is irrelevant
at $h\gtrsim l_2$, leading to the same orientation field above the substrate blocks. 
On the other hand, the value of $\alpha_\infty(\infty;l_1/l_2)$  decreases as $l_1/l_2$ increases because the 
final anchoring results from a competition between the homeotropic anchoring favoured by the top of the blocks, and 
the planar anchoring favoured by the rectangular wells.  

The equilibrium texture for each substrate is that which minimizes the free energy. 
First, we note that the leading-order contributions due to the cusp singularities
Eq. (\ref{asymptotic-core}) are equal to $2\pi K/3$ for all the textures. Therefore, 
this term will be irrelevant to identify
which texture is the equilibrium one for a given substrate relief, so we need to analyse the next-to-leading order contributions.
As shown in Ref. \cite{oscar} for the sawtooth substrate, we have to consider a term of elastic origin, 
in addition to the contribution of the cores corresponding to the disclination-like singularities close to the cusps,
to fully account for the next-to-leading contribution to the interfacial free energy per $z-$unit length 
and $x-$period. 
First we analyse the elastic contribution,
which depends on $l_1$, $l_2$ and $h$ through two independent ratios $h/l_2$ and $l_1/l_2$, or equivalently, 
on the roughness $r=1+2h/(l_1+l_2)$ and $l_1/l_2$ \cite{patricio5}. 
The results of our calculations show that the $N^o_2$ texture has always a higher elastic free
energy than the other textures, so it can be discarded from the discussion. 
Fig. \ref{fig7} illustrates this point showing that the distortions of the nematic director field are more 
pronounced in the $N^o_2$ texture than in the other textures.
Another interesting observation is that both symmetric textures have the same elastic free energy. 
This is shown analytically in the \ref{appendixc}, where the exact elastic contribution to the free energy of the symmetric 
textures is calculated. The numerical results are in
excellent agreement with the analytical results, as can be seen in Fig. \ref{fig9}(a), although some
deviations are visible for very shallow and/or narrow crenels. This observation 
provides a stringent test of the numerical accuracy. Purely elastic arguments 
predict that the $N^o_1$ state is the lowest free-energy texture at all crenellated substrates, as shown in Fig.
\ref{fig9}(a). However, the free-energy of the symmetric and asymmetric textures approach each other
at small values of the roughness and thus, the cusp singularity cores contribution to the free energy
may stabilize the symmetric textures with respect to the tilted one.  
In Fig. \ref{fig9}(b) we plot the core contributions associated to the different cusps and
topological charges, as well as the total contribution for each nematic texture. 
Note that the total core contribution for a surface state corresponding to a nematic texture
is just the sum of the contributions associated to each isolated cusp, regardless the substrate geometry. This 
contribution breaks the free-energy degeneracy of the symmetric textures, favouring the 
$N^\perp_2$ texture at small and large values of $w$, and the $N^\perp_1$ texture otherwise. 
The core contribution associated to the $N^o_1$ texture is always higher
than that corresponding to the least free-energy symmetric texture, since the tilted texture core contribution is 
the average of the values of the symmetric textures. So, if the core contribution of the tilted configuration exceeds
the elastic free-energy difference between the symmetric and the $N_1^o$ textures, the corresponding 
symmetric state may be stabilized. Fig. \ref{fig10} depicts the global phase diagram of the crenellated substrate.
At large substrate roughness, the tilted nematic texture is the most
stable phase. By decreasing the roughness, a transition to a symmetric texture 
may be observed. These findings are in agreement with previous experimental \cite{luo} and Landau-de Gennes numerical
\cite{patricio5} results. At low and high values of the anchoring parameter $w$, 
the symmetric state is $N^\perp_2$, while for intermediate values of $w$
it is $N^\perp_1$. Furthermore, reentrant behaviour is found at intermediate values of $w$.
The phase boundaries move to higher values of $h/l_2$ as $l_1/l_2$, but
they saturate at $h/l_2\gtrsim 1$. 

Finally, we comment on the effect of the anisotropy of the elastic 
constants. As discussed for the sawtooth substrate, the main effect of the anisotropy 
is to shift the leading-order elastic free-energy contribution. Therefore, if
$\lambda/\xi_0$ is large, then the $N^\perp_1$ ($N^\perp_2$) texture is favoured when 
$K_3>K_1$ ($K_3<K_1$), respectively.
By comparison with the sawtooth substrate, the leading contributions are again identical for the three nematic textures
at crenellated substrate where blocks have tilted lateral sides. 
However, if $|K_3/K_1-1|\ln(\lambda/\xi_0)$ is of order of the next-to-leading 
contribution when $K_1=K_3$, then this is another contribution to take into
account when evaluating the phase diagram.  

\subsection{The sinusoidal substrate}

We now turn to a sinusoidal substrate of period $\lambda$ and amplitude 
$A$. This has also been studied previously \cite{berreman,Fukuda,barbero,patricio4,raisch}. 
As the substrate relief does not have cusps, a state without defects is expected to be the least free-energy 
state. This state exhibits homeotropic anchoring, i.e. the far-field
nematic director is oriented along the $y$ axis, for all $qA$ (see
Fig. \ref{fig11}(a)). We denote this texture by $N^\perp$. 
As $qA$ increases, the substrate roughness
increases, and the orientational field exhibits large distorsions to follow
the anchoring at the substrate. Numerical results show that, under
these circumstances, the elastic distortions are lowered by  
reorienting the nematic director field, in the groove, along the $x$
axis (see Fig. \ref{fig11}(b)), and thus the texture exhibits planar 
anchoring, denoted by $N^\parallel$. 
This texture involves the nucleation of two disclination lines with opposite 
topological charges, located by symmetry above the top and bottom of the 
substrate relief, at a distance plotted in Fig. \ref{fig12}. This distance is proportional to 
$\lambda$, and in the limit of large $\lambda$, the disclination lines are not
affected by the substrate. The distance decreases as the substrate
roughness increases until the it stabilizes for $qA>2$. Thus, in an
effective way, the
disclinations lines are bound to the surface relief (on the $\lambda$
scale), driving the orientational field almost horizontal 
everywhere.
At large $\lambda$, the interfacial free energy of the 
$N^\perp$ texture depends on $A$
and $\lambda$ through the factor $qA$ which determines the substrate roughness.
On the other hand, from Eq. (\ref{asymptotic-core}) 
the interfacial free energy of the $N^\parallel$ texture
has a leading contribution $(K\pi/2) \ln \lambda/\xi_0$, and the 
next-to-leading term has the same $qA$ dependence as above. Note that, in
this case, we have to add the core contributions associated to the $\pm 1/2$
disclinations lines, with constant values $f_c(I=+1/2)/K=0.63\pm 0.01$ and 
$f_c(I=-1/2)/K=-0.14\pm 0.01$. Thus, at large 
$\lambda$ only the $N^\perp$ texture is expected for any substrate
roughness. However, the weak dependence of the interfacial
free energy of the $N^\parallel$ texture on $\lambda$ implies that, for moderate values
of $\lambda$, an anchoring transition between the $N^\perp$ and $N^\parallel$
textures may be observed. Fig. \ref{fig13} shows the interfacial free 
energy of the $N^\perp$ and $N^\parallel$ textures 
as a function of $qA$, for
different values of $\lambda$.  While the $N^\perp$ branch depends only 
on $qA$ and is an increasing function of this parameter, the $N^\parallel$
branches are decreasing functions of $qA$, and for different values
of $\lambda$ are shifted by the $\ln\lambda$ term.

Fig. \ref{fig14} shows the $qA-\lambda$ anchoring phase diagram. The 
homeotropic anchoring state $N^\perp$ is favoured at low $qA$, while at large
substrate roughness planar anchoring is observed, i.e. the $N^\parallel$ 
has the lowest free energy. We note that the value of 
$\lambda$ at the transition increases almost exponentially with $qA$.  

Finally, as in the previous cases we can include the effect
of the anisotropy of the elastic constants perturbatively. If we assume that the elastic 
distortions are on the $xy$ plane, the conclusion
is that the anchoring transition, which corresponds to moderate values
of $\lambda$, may be shifted by this contribution, although qualitatively 
it will be very similar. However, if the twist elastic constant is smaller than 
$K$, there is experimental \cite{ohzono} and numerical \cite{nuno_submitted} evidence 
of a twist instability which breaks the azimuthal symmetry: the disclination line is no
longer parallel to the $z$ axis, but exhibits a zig-zag structure which decreases the splay and
bending distortions. This cannot happen in the sawtooth and crenellated substrates, since 
the disclination-like singularities are located at the surface cusps.

\section{Conclusions \label{sec5}}

In this paper we report the results of a numerical investigation of the equilibrium nematic textures at patterned 
substrates under strong anchoring conditions. We characterize the surface phase diagram of nematic
textures which differ in the tilt angle of the far-field nematic director with respect to the substrate reference plane.
First-order phase transitions between these surface states, i.e. anchoring transitions, are observed when the geometric 
features of the surface relief are varied, although there are other control parameters (such as the anchoring surface 
strength) which may play a role in
the location of the phase boundaries. Our findings, which generalize previous work by the authors \cite{oscar}, differ from 
previous results for weak anchoring conditions, where
these anchoring transitions are driven by the competition between the elastic deformations in the nematic orientational
field and the surface anchoring energy. By contrast, in the strong-anchoring regime, these transitions are a direct 
outcome of the interplay between the elastic deformations and the formation of disclinations 
and disclination-like singularities near surface cusps. In addition, a small elastic anisotropy can play a similar 
role to that of these topological defects. To illustrate our study, we consider
three substrate reliefs: the sawtooth substrate, the crenellated substrate and the sinusoidal subtrate. 
For the sawtooth and sinusoidal substrates, we observed a homeotropic to planar anchoring transition as the 
substrate roughness is increased. On the other hand, the crenellated substrate exhibits a more complex anchoring phase
diagram, with a homeotropic to oblique anchoring transition, which depends not only on the substrate roughness but also 
on the surface anchoring strength. The latter results from the dependence of the core contribution of the cusp singularities on the anchoring strength.

Some final remarks are in order.
Although we used the Landau-de Gennes model to obtain the defect core contributions
to the free energy, any other model could be considered. This may change the results quantitatively, when this 
contribution is relevant as for the crenellated substrates, but not qualitatively. 
Secondly, our procedure can be easily modified to 
consider the presence of nematic-isotropic interfaces. This allows the study of wetting, filling and related interfacial
phenomena for nematic liquid crystals. This is ongoing work, and will be published elsewhere. Finally, we restricted the nematic orientational distortions to the plane perpendicular to the longitudinal axis of the surface. The generalization to full three-dimensional systems to consider situations where twist \cite{ohzono,jeong} or saddle-splay 
\cite{davidson2,nayani} distorsions play a role is a formidable task which is currently beyond the scope of our work.

\ack
We acknowledge financial support from the Portuguese Foundation for Science and Technology under Contracts Nos. EXCL/FIS-NAN/0083/2012 and UID/FIS/00618/2013 (NMS and MMTG).
O.A.R.-G. and J.M.R.-E. also acknowledge partial financial support from the Spanish Ministerio de Econom\'{\i}a y Competitividad through
grant no. FIS2012-32455, and Junta de Andaluc\'{\i}a through grant no. P09-FQM-4938, all co-funded
by the EU FEDER.

\appendix
\section{Evaluation of $\theta_{ns}$ using the boundary element method\label{appendixa}}
The field $\theta_{ns}$, as a solution of the Laplace equation on $R$, has the boundary integral representation:
\begin{eqnarray} 
&&\theta_{ns}(\mathbf{r})=\oint_{\partial R} d \mathbf{s} \Bigg(
[\boldsymbol{\nu}(\mathbf{s})\cdot \boldsymbol{\nabla}_{\mathbf{s}} 
\theta_{ns}(\mathbf{s})]
G(\mathbf{s},\mathbf{r})
\label{bem1}\\
&&-\theta_{ns}(\mathbf{s})
\left[\boldsymbol{\nu}(\mathbf{s})\cdot \boldsymbol{\nabla}_{\mathbf{s}}
G(\mathbf{s},\mathbf{r})\right]
\Bigg)
\nonumber
\end{eqnarray}
where the contour integral over the boundary $\partial R$ of $R$ is
counter-clockwise, $\boldsymbol{\nu}(\mathbf{s})$ is the outwards normal
to the boundary at $\mathbf{s}$ and $G(\mathbf{r},\mathbf{r}_0)$ is the
fundamental solution of the Laplace equation in the infinite strip $-\lambda/2\le x\le \lambda/2$, $-\infty<y<\infty$ 
with periodic boundary conditions on $x$:
\begin{equation}
G(\mathbf{r},\mathbf{r}_0)=-\frac{1}{4\pi}\ln\left(\cosh q(y-y_0)-\cos q(x-x_0)\right)
\end{equation}
where $\mathbf{r}=(x,y)$ and $\mathbf{r}_0=(x_0,y_0)$. Note that this solution can be obtained 
as the composition of the fundamental solution on the free plane $-\ln |\mathbf{r}|/(2\pi)$
and the conformal mapping $\zeta=\sin q(z-z_0)/2$ which maps the strip onto the full complex plane.
As both $\theta_{ns}$ and $G$ are periodic on $x$ with period $\lambda$, the contributions to the integral
(\ref{bem1}) from the lateral sides $x=\pm \lambda/2$ cancel each other. On the other hand, 
$G(\mathbf{r},\mathbf{r}_0)\approx -|y-y_0|/(2\lambda)+\ln 2/(4\pi)$ at large $y$. As we impose $\theta_{ns}\to 
\theta_{ns}^\infty$ and $\partial \theta_{ns}/\partial y\to 0$ as $y\to \infty$, the contribution to the
integral (\ref{bem1}) from the top boundary $y=H\to \infty$ is equal to $\theta_{ns}^\infty/2$.
Therefore, Eq. (\ref{bem1}) can be rewritten as:
\begin{eqnarray} 
&&\theta_{ns}(\mathbf{r})=\frac{\theta_{ns}^\infty}{2}+\int_{\cal L} d \mathbf{s} \Bigg( 
[\boldsymbol{\nu}(\mathbf{s})\cdot \boldsymbol{\nabla}_{\mathbf{s}} 
\theta_{ns}(\mathbf{s})]
G(\mathbf{s},\mathbf{r})
\label{bem1-2}\\
&&-\theta_{ns}(\mathbf{s})
\left[\boldsymbol{\nu}(\mathbf{s})\cdot \boldsymbol{\nabla}_{\mathbf{s}}
G(\mathbf{s},\mathbf{r})\right]\Bigg)
\nonumber
\end{eqnarray}
We impose Dirichlet boundary conditions on the substrate relief, so the last term on the right-hand
side of Eq. (\ref{bem1}) is known. The unknowns are the normal derivatives of
$\theta_{ns}$ on the substrate and the far-field value $\theta_{ns}^\infty$. The former is obtained by solving the 
integral equation \cite{brebbia,katsikadelis}:
\begin{eqnarray} 
&&\int_{\cal L} d \mathbf{s} 
[\boldsymbol{\nu}(\mathbf{s})\cdot \boldsymbol{\nabla}_{\mathbf{s}} 
\theta_{ns}(\mathbf{s})]
G(\mathbf{s},\mathbf{s}_0)=\frac{\theta_{ns}(\mathbf{s}_0)-\theta_{ns}^\infty}{2}
\label{bem2}
\\
&&+\int_{\cal L} d \mathbf{s} \theta_{ns}(\mathbf{s})
\left[\boldsymbol{\nu}(\mathbf{s})\cdot \boldsymbol{\nabla}_{\mathbf{s}}
G(\mathbf{s},\mathbf{s}_0)\right]
\nonumber
\end{eqnarray}
where $\mathbf{s}_0\in {\cal L}$. On the other hand, for large $y$, $\theta_{ns}(\mathbf{r})\approx 
\theta_{ns}^{\infty}$ and Eq. (\ref{bem1-2}) reduces to
\begin{eqnarray}
\theta_{ns}^{\infty}&=&\int_{\cal L} d \mathbf{s}  
[\boldsymbol{\nu}(\mathbf{s})\cdot \boldsymbol{\nabla}_{\mathbf{s}} 
\theta_{ns}(\mathbf{s})]\left(\frac{s_y-y}{\lambda}+\frac{\ln 2}{2\pi}\right)\nonumber\\
&-&\frac{1}{\lambda}\int_{\cal L} d \mathbf{s}\theta_{ns}(\mathbf{s})\nu_y(\mathbf{s}) 
\nonumber\\
&=&\frac{1}{\lambda}\int_{\cal L} d \mathbf{s}\left( s_y [\boldsymbol{\nu}(\mathbf{s})\cdot 
\boldsymbol{\nabla}_{\mathbf{s}} 
\theta_{ns}(\mathbf{s})] - \theta_{ns}(\mathbf{s})\nu_y(\mathbf{s})\right)
\label{bem3}
\end{eqnarray}
where $s_y$ and $\nu_y$ are the $y$-components of $\mathbf{s}$ and $\boldsymbol{\nu}$, respectively,
and the second equality results from the fact that 
\begin{eqnarray}
0&=&\int_{\cal A}d\mathbf{r}\nabla^2 \theta_{ns}= \oint d \mathbf{s} [\boldsymbol{\nu}(\mathbf{s})\cdot 
\boldsymbol{\nabla}_{\mathbf{s}} 
\theta_{ns}(\mathbf{s})]
\label{condition_integral}\\
&=&\int_{\cal L} d \mathbf{s} [\boldsymbol{\nu}(\mathbf{s})\cdot 
\boldsymbol{\nabla}_{\mathbf{s}} 
\theta_{ns}(\mathbf{s})]
\nonumber
\end{eqnarray}
Finally, substituting Eq. (\ref{bem3}) in Eq. (\ref{bem2}), we obtain:
\begin{eqnarray} 
&&\int_{\cal L} d \mathbf{s} 
[\boldsymbol{\nu}(\mathbf{s})\cdot \boldsymbol{\nabla}_{\mathbf{s}} 
\theta_{ns}(\mathbf{s})]
\left(G(\mathbf{s},\mathbf{s}_0)+\frac{s_y}{2\lambda}\right)=\frac{\theta_{ns}(\mathbf{s}_0)}{2}
\label{bem4}
\\
&&+\int_{\cal L} d \mathbf{s} \theta_{ns}(\mathbf{s})
\left[\boldsymbol{\nu}(\mathbf{s})\cdot \boldsymbol{\nabla}_{\mathbf{s}}
G(\mathbf{s},\mathbf{s}_0)+\frac{\nu_y}{2\lambda}\right]
\nonumber
\end{eqnarray}
In order to solve Eq. (\ref{bem4}), we
discretize the boundary as a set of straight segments (the boundary elements).
It is important to ensure that the substrate cusps correspond to extremes of these segments.
We use the constant boundary element approach \cite{katsikadelis}, and thus 
assume that both $\theta_{ns}$ and its normal derivative are constant along
each boundary element. Introducing this approximation into Eq. (\ref{bem4}),
we obtain a set of linear algebraic equations for the normal derivatives.
Once this is solved, the far-field orientation $\theta_{ns}^\infty$ is 
obtained from Eq. (\ref{bem3}).
\section{Evaluation of the elastic contribution to the interfacial free energy $f_e$ \label{appendixb}}
The value of $f_e$ can be expressed as a contour integral, Eq. (\ref{contourfreeenergy}). 
Using the periodicity of $\theta$ on the boundaries $x=\pm\lambda/2$ and the free
boundary at $y\to \infty$, the contour integral on the right-hand side of Eq. (\ref{contourfreeenergy}) is written as:
\begin{equation}
\frac{K}{2}\int_{\cal L}\theta (\boldsymbol{\nu\cdot\nabla}\theta) d\mathbf{s}+
\sum_{j=1}^{N_D} 2\pi I_j \int_{y_j}^\infty dy\left(\frac{\partial \theta}{\partial x}\right)_{x=x_j} 
\label{energy2}
\end{equation}
where the first integral is on the surface relief, and the other terms are on the branch cuts ${\cal B}_j$
starting at the disclination $j$ position $(x_j,y_j)$ with topological charge $I_j$. Finally, we use  
$\theta=\theta_s+\theta_{ns}$ in the derivatives, and the elastic contribution becomes:
\begin{eqnarray}
\frac{K}{2}\int_{\cal L}\theta (\boldsymbol{\nu\cdot\nabla}\theta_s) d\mathbf{s}+
\sum_{j=1}^{N_D} 2\pi I_j \int_{y_j}^\infty dy\left(\frac{\partial \theta_s}{\partial x}\right)_{x=x_j}
\label{energy3}\\
+\frac{K}{2}\int_{\cal L}\theta (\boldsymbol{\nu\cdot\nabla}\theta_{ns}) d\mathbf{s}
+\sum_{j=1}^{N_D} 2\pi I_j \int_{y_j}^\infty dy\left(\frac{\partial \theta_{ns}}{\partial x}\right)_{x=x_j}
\nonumber
\end{eqnarray}
The first two terms exhibit singularities associated to the cusps (first integral) and disclination cores (second term) 
which must be handled carefully by deforming the contour with arcs of circle of radii $\xi_0$ to avoid them. 
In fact, these singularities lead to the contribution Eq. (\ref{asymptotic-core}) mentioned above.
The first integral can be obtained using the substrate relief discretization 
considered to obtain $\theta_{ns}$. Thus, if we consider the boundary as the union of ${\cal L}_k$ segments 
($k=1,\ldots,N_{e}$, ordered counterclockwise), the first 
integral may be approximated as:
\begin{equation}
\frac{K}{2}\int_{\cal L}\theta (\boldsymbol{\nu\cdot\nabla}\theta_s) d\mathbf{s}\approx
\sum_{k=1}^{N_{e}} \frac{K}{2} \theta_k 
\boldsymbol{\nu}_k\boldsymbol{\cdot}\int_{{\cal L}_k}\boldsymbol{\nabla}\theta_s 
d\mathbf{s}
\label{energy4}
\end{equation}
where $\theta_k$ and $\boldsymbol{\nu}_k$ are the segment midpoint and the outwards unit normal to that segment 
${\cal L}_k$, respectively. Substitution of the expression for $\theta_s$ Eq. (\ref{thetas}) into Eq. (\ref{energy4}) leads
after some algebra to:
\begin{eqnarray}
\frac{K}{2}\sum_i I_i^2 \Delta \phi_i\left(\ln \left(\frac{\lambda}{\xi_0}\right)
-\ln \sqrt{2}\pi+\frac{qy_i}{2}
\right)
\label{energy5}\\
+ 
\frac{K}{2}\sum_i \sum_{k}^\prime I_i \Delta \theta_k \nonumber\\
\times\Bigg(\frac{1}{2}\ln\left(\cosh q(y_k-y_i)-\cos q(x_k-x_i)\right)-
\frac{q}{2}y_k\Bigg)\nonumber\\
+\frac{K}{2}\sum_j \sum_{k} I_j \Delta \theta_k \nonumber\\
\times\Bigg(\frac{1}{2}\ln\left(\cosh q(y_k-y_j)-\cos q(x_k-x_j)\right)+
\frac{q}{2}y_k\Bigg)\nonumber
\end{eqnarray} 
The first term in Eq. (\ref{energy5}) corresponds to the contribution associated to the surface
cusps. In the second term, the sum on $i$ is over the number of cusps, while the sum on $k$ is
over the boundary elements. In this expression, $(x_i,y_i)$ is the position of the cusp, $I_i$ is the
effective topological charge of the cusp singularity, 
$(x_k,y_k)$ are the coordinates of the left extreme of the segment ${\cal L}_k$ and
$\Delta \theta_k=\theta_k-\theta_{k-1}$ with $\theta_0\equiv \theta_{N_{e}}$. The prime denotes
that we exclude from the sum the nodes that correspond to surface cusps. 
Finally, in the last term the sum on $j$ is over the disclination lines at positions $(x_j,y_j)$ with topological
charges $I_j$, while the other terms have the same meaning
as before. 

The second term in Eq. (\ref{energy3}) can be obtained analytically as:
\begin{eqnarray}
\frac{K\pi N_D}{4}\left(\ln\left(\frac{\lambda}{\xi_0}\right)-\ln\pi\right)+\frac{K\pi}{4}\sum_j qy_j\label{energy6}\\
+\frac{K}{2}\sum_i \sum_j 2\pi I_i I_j \Bigg(\frac{q}{2}(y_i-y_j)+\frac{\ln 2}{2}\nonumber\\ 
+\frac{1}{2}\ln
\left(\cosh q(y_i-y_j)-\cos q(x_i-x_j)\right)\Bigg)\nonumber\\
+\frac{K}{2}\sum_j \sum_{j'}^\prime 2\pi I_j I_{j'} 
\Bigg(\frac{q}{2}(y_{j'}+y_j)+\frac{\ln 2}{2}\nonumber\\
+\frac{1}{2}\ln
\left(\cosh q(y_{j'}-y_j)-\cos q(x_{j'}-x_j)\right)\Bigg)\nonumber 
\end{eqnarray} 
where the sum on $i$ is over the surface cusps, the sums on $j$ and $j'$ are over the disclination lines, and
in the last term the prime denotes that we exclude from the sum terms with $j=j'$.

The last two contributions to Eq. (\ref{energy3}) associated with $\theta_{ns}$ can be obtained in a similar way.
The third term yields
\begin{equation}
\frac{K}{2}\int_{\cal L}\theta (\boldsymbol{\nu\cdot\nabla}\theta_{ns}) d\mathbf{s}
\approx \sum_{k=1}^{N_{e}} \frac{K}{2} \ell_k \theta_k (\boldsymbol{\nu\cdot\nabla}\theta_{ns})_k
\label{energy7}
\end{equation}
where $\ell_k$ and $\theta_k$ are the length and the value of $\theta$ at the midpoint of ${\cal L}_k$, respectively. 
On the other hand, $(\boldsymbol{\nu\cdot\nabla}\theta_{ns})_k$
is the value of the non-singular normal derivative obtained from Eq. (\ref{bem4}). Finally, the last term in 
Eq. (\ref{energy3}) can be obtained by standard integration techniques, where $\theta_{ns}$ in the branch cut
is evaluated using Eq. (\ref{bem1-2}).

\section{Exact elastic contribution to the free energy density for the symmetric textures on crenellated substrates\label{appendixc}}

In this Appendix we will evaluate the elastic free energy of the symmetric textures on
a crenellated substrate characterized by a block of width $l_1$ and height $h$, and period $\lambda=l_1+l_2$.
We note that, due to the symmetry of the texture, we may evaluate the elastic free energy 
in the domain shown in Fig. \ref{fig_appendix}(a), which corresponds to half a period of the substrate
relief. The domain boundary follows the substrate except close to its cusps, where it is rounded by
arcs of circle of radii $\xi_0$. The vertical lateral sides run from the subtrate to infinity. 
We set the origin on the lower substrate cusp, so the upper substrate cusp is at 
$(0,h)$. On the domain boundary we set $\theta=0$ except for the vertical segment which joins the two
surface cusps, where we set either $\theta=\pi/2$ or $\theta=-\pi/2$ (depending on the symmetric
texture considered).    

We map the domain in the $z$-plane to the upper half $\zeta$-plane by using the following Schwarz-Christoffel
transformation:
\begin{eqnarray}
z=C\int \frac{\sqrt{\zeta+1}d\zeta}{\sqrt{\zeta+a}\sqrt{\zeta}\sqrt{\zeta-b}}+C'
\label{SCtransform}\\
=C'+\frac{2C}{\sqrt{a+b}}\Bigg[(1+b)F\left(\gamma\Bigg|\frac{a(1+b)}{a+b}\right)
\label{SCtransform2}\\
-b\Pi\left(\frac{a}{a+b};\gamma\Bigg|\frac{a(1+b)}{a+b}\right)\Bigg]
\nonumber\\
\textrm{with }\gamma=\arcsin\left[\textrm{sign}\left(\frac{\zeta}{\zeta-b}\right)
\sqrt{\frac{(a+b)\zeta}{a(\zeta-b)}}\right]
\nonumber
\end{eqnarray}
where $C$ and $C'$ are complex constants, $a>1$ and $b>0$ real numbers, and $F(z|m)$ and 
$\Pi(n;z|m)$ are the incomplete elliptic integral of the first and third kind, respectively.
The points $-l_1/2+ih$, $ih$, $0$ and $l_2/2$ in the $z-$plane are mapped onto
$-a$, $-1$, $0$ and $b$, respectively. These conditions fix the values of $C$, $C'$, $a$ and $b$. 
In particular, as $F(0|m)=\Pi(n;0|m)=0$, we find that $C'=0$. On the other hand, the real part of
Eq. (\ref{SCtransform2}) diverges as $\zeta\to \pm \infty$, but the imaginary parts have well defined
values which differ by $\pi C$. Consequently, $C=i(l_1+l_2)/2\pi$.
Finally, the values of $a$ and $b$ are determined by the requirement that the image of the point
$-l_1/2+ih$ is $-a$ in the $\zeta$-plane. This leads to the conditions:
\begin{eqnarray}
\frac{h}{l_1+l_2}=\frac{1}{\pi\sqrt{a+b}}\Re\Bigg[(1+b)K\left(\frac{a(1+b)}{a+b}\right)
\label{SCtransform3}\\
-b\Pi\left(\frac{a}{a+b}\Bigg|\frac{a(1+b)}{a+b}\right)\Bigg]
\nonumber\\
\frac{l_1}{l_1+l_2}=\frac{2}{\pi\sqrt{a+b}}\Im\Bigg[(1+b)K\left(\frac{a(1+b)}{a+b}\right)
\label{SCtransform4}\\
-b\Pi\left(\frac{a}{a+b}\Bigg|\frac{a(1+b)}{a+b}\right)\Bigg]
\nonumber
\end{eqnarray} 
where $\Re(z)$ and $\Im(z)$ are the real and imaginary parts of $z$, respectively,
and $K(m)$ and $\Pi(n|m)$ are the complete elliptic integrals of the first and third kinds, respectively.
The values of $a$ and $b$ must be obtained numerically from these conditions for given values
of $h/(l_1+l_2)$ and $l_1/(l_1+l_2)$.

In order to obtain the orientational field $\theta$, we solve the Laplace equation in the upper 
half $\zeta$-plane, with the real axis as boundary, except close to 
$-1$ and $0$, where it is rounded by arcs of circle of radii $\epsilon_1$ and $\epsilon_2$, respectively.  
The boundary conditions are $\tilde\theta=0$ for $\zeta<-1-\epsilon_1$ and $\zeta>\epsilon_2$, and 
$\pm \pi/2$ for $-1+\epsilon_1\le z \le -\epsilon_2$. The solution is:
\begin{equation}
\tilde{\theta}(\zeta)=\pm\frac{1}{2}\left[\arctan\left(\frac{\tilde y}{\tilde x}\right)-
\arctan\left(\frac{\tilde y}{\tilde{x}+1}\right)\right]
\label{solution}
\end{equation}
where $\zeta=\tilde{x}+i\tilde{y}$. The orientational field in the original domain is $\theta(x,y)=
\tilde{\theta}(\zeta(z))$, where $z=x+iy$.
Thus the free energy per $z-$unit length and $x-$period is
\begin{eqnarray}
&&K\int d x d y 
|\boldsymbol \nabla \theta|^2=K\int d \tilde{x} d \tilde{y}
|\boldsymbol \nabla' \tilde \theta|^2 
\nonumber\\
&&=\mp\frac{K\pi}{2} 
\int_{-1+\epsilon_1}^{-\epsilon_2} \left(
\frac{\partial \tilde \theta}{\partial \tilde y}\right) d\tilde{x} = -\frac{K\pi}{4}\ln(\epsilon_1\epsilon_2)
\label{mfofreeenergy}
\end{eqnarray}
We can use Eq. (\ref{SCtransform}) to relate $\epsilon_1$ and $\epsilon_2$ to $\xi_0$ \cite{oscar}, 
leading to:
\begin{eqnarray}
\epsilon_1=\left(3\pi\frac{\xi_0}{l_1+l_2}\sqrt{a-1}\sqrt{b+1}\right)^{2/3}\label{epsilon1}\\
\epsilon_2=\left(\pi
\frac{\xi_0}{l_1+l_2}\sqrt{ab}\right)^2
\label{epsilon2}
\end{eqnarray}
Therefore, the free energy Eq. (\ref{mfofreeenergy}) can be recast as:
\begin{equation}
\frac{2\pi K}{3}\ln \frac{l_1+l_2}{\xi_0} -\frac{K\pi}{6} \ln \left(3\pi^4 \sqrt{a-1}\sqrt{b+1}(ab)^{3/2}
\right)
\label{mfofreeenergy2}
\end{equation}
Note that this expression depends only on the geometric characteristics of the surface relief, and not 
on the boundary condition which determines the symmetric nematic texture. Thus, both symmetric textures have
exactly the same elastic contribution to the free energy. This is consistent with the results for the single
step solution \cite{davidson}, which would correspond to our case in the limit $l_1\to \infty$ and $l_2\to \infty$.

\clearpage
\renewcommand{\thefigure}{\arabic{figure}}
\setcounter{figure}{0}
\vspace*{0.5cm}
\begin{figure}[H]
\centerline{\includegraphics[width=0.9\columnwidth]{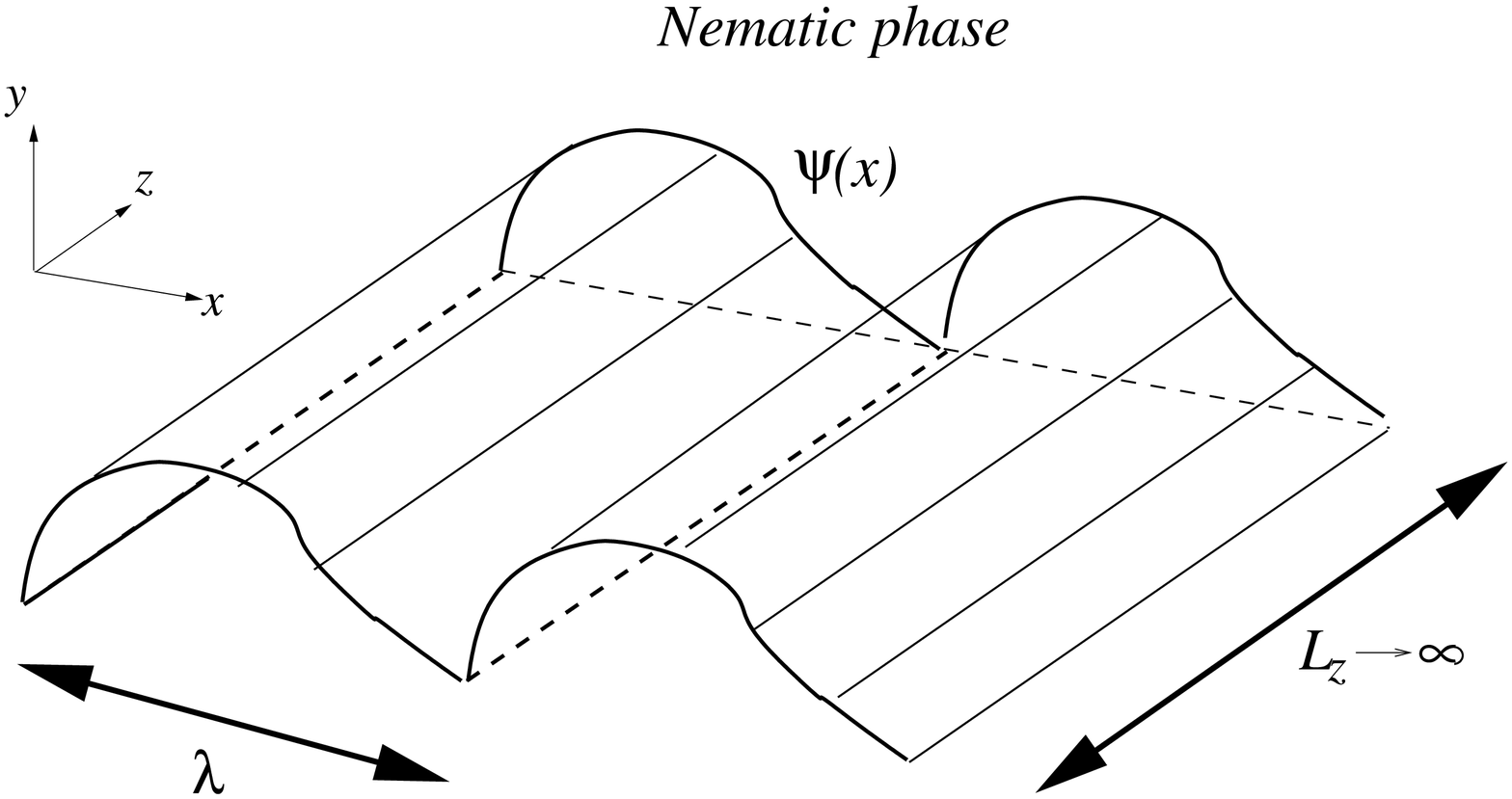}}
\caption
{Schematic picture of the substrate geometry, characterized by the
wavenumber $\lambda$ and the relief profile $\psi(x)$.
}
\label{fig1}
\end{figure}
\clearpage
\vspace*{0.5cm}
\begin{figure}[H]
\centerline{\includegraphics[width=0.9\columnwidth]{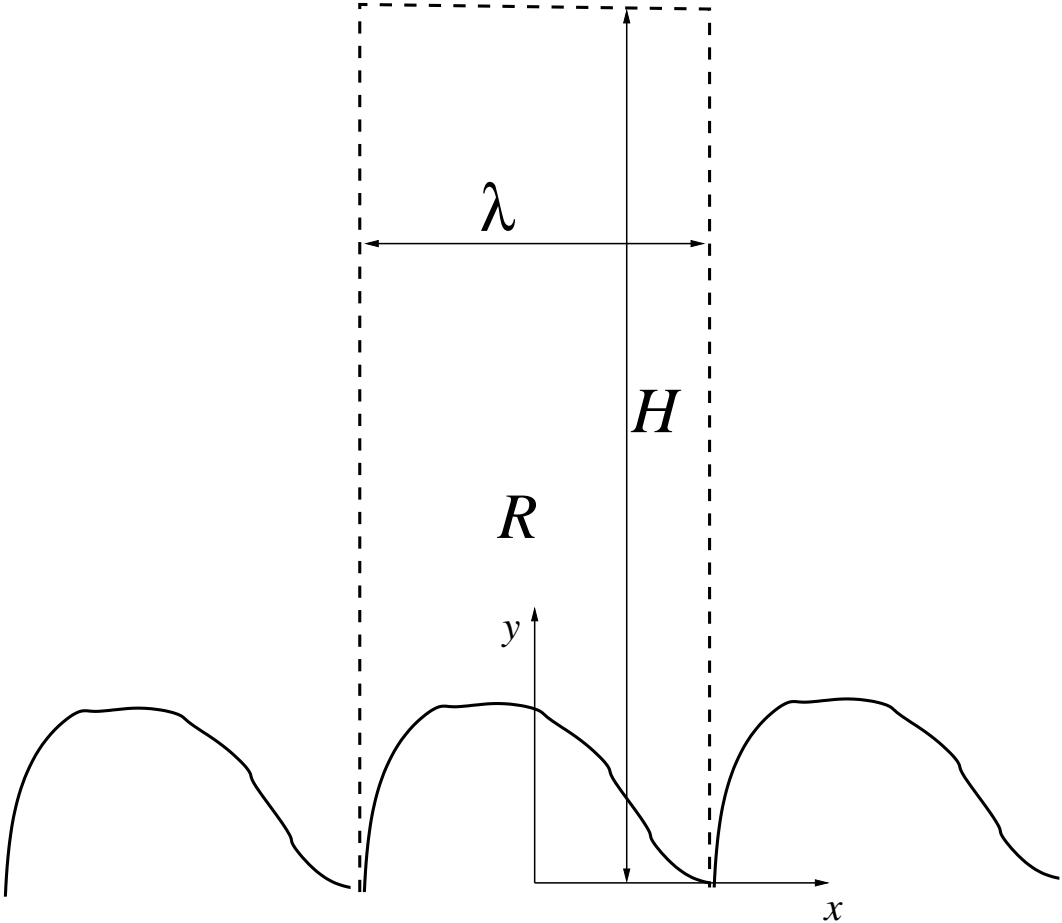}}
\caption
{Plot of the region $R$ where the equation for the orientational field $\theta$ is solved. $\lambda$ is the
period of the substrate relief and $H$ is the height of the domain.}
\label{fig2}
\end{figure}
\clearpage
\vspace*{0.5cm}
\begin{figure}[H]
\centerline{\includegraphics[width=0.8\columnwidth]{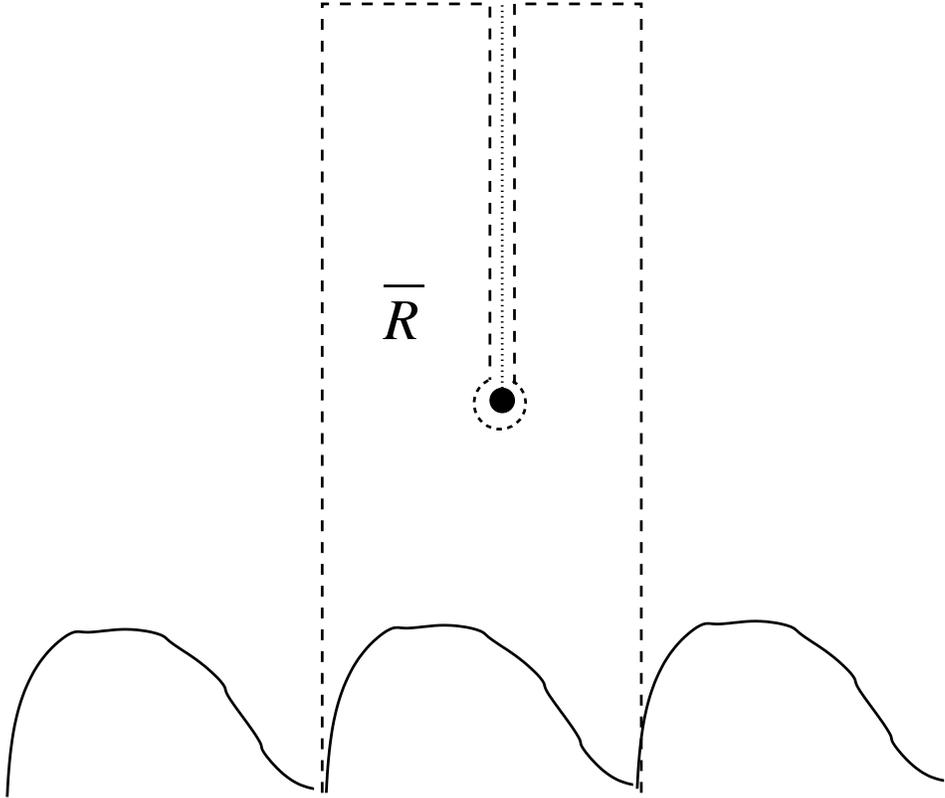}}
\caption
{Plot of the modified region $\overline{R}$ for a nematic with a $\pm 1/2$-disclination 
line in the bulk, represented by the filled circle. The vertical dotted line corresponds 
to the branch cut, where a discontinuity of $\pm \pi$ in the orientational field is 
observed. The dashed line corresponds to the boundary of $\overline{R}$.} 
\label{fig3}
\end{figure}
\clearpage
\vspace*{0.5cm}
\begin{figure}[H]
\centerline{\includegraphics[width=0.8\columnwidth]{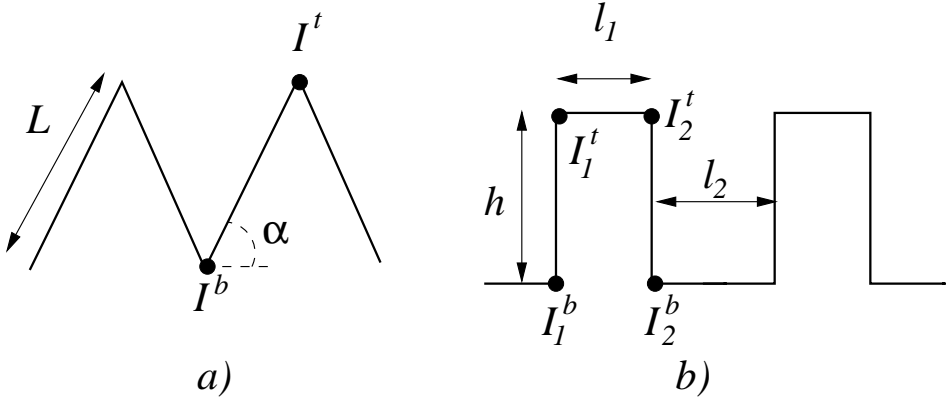}}
\caption
{Geometric characteristics of: (a) a sawtooth substrate; (b) a crenellated substrate. The positions and effective 
topological charges $I$ associated to the cusps of one substrate relief period are highlighted.}
\label{fig4}
\end{figure}
\clearpage
\vspace*{0.5cm}
\begin{figure}[H]
\begin{tabular}{cc}
\includegraphics[width=0.45\columnwidth]{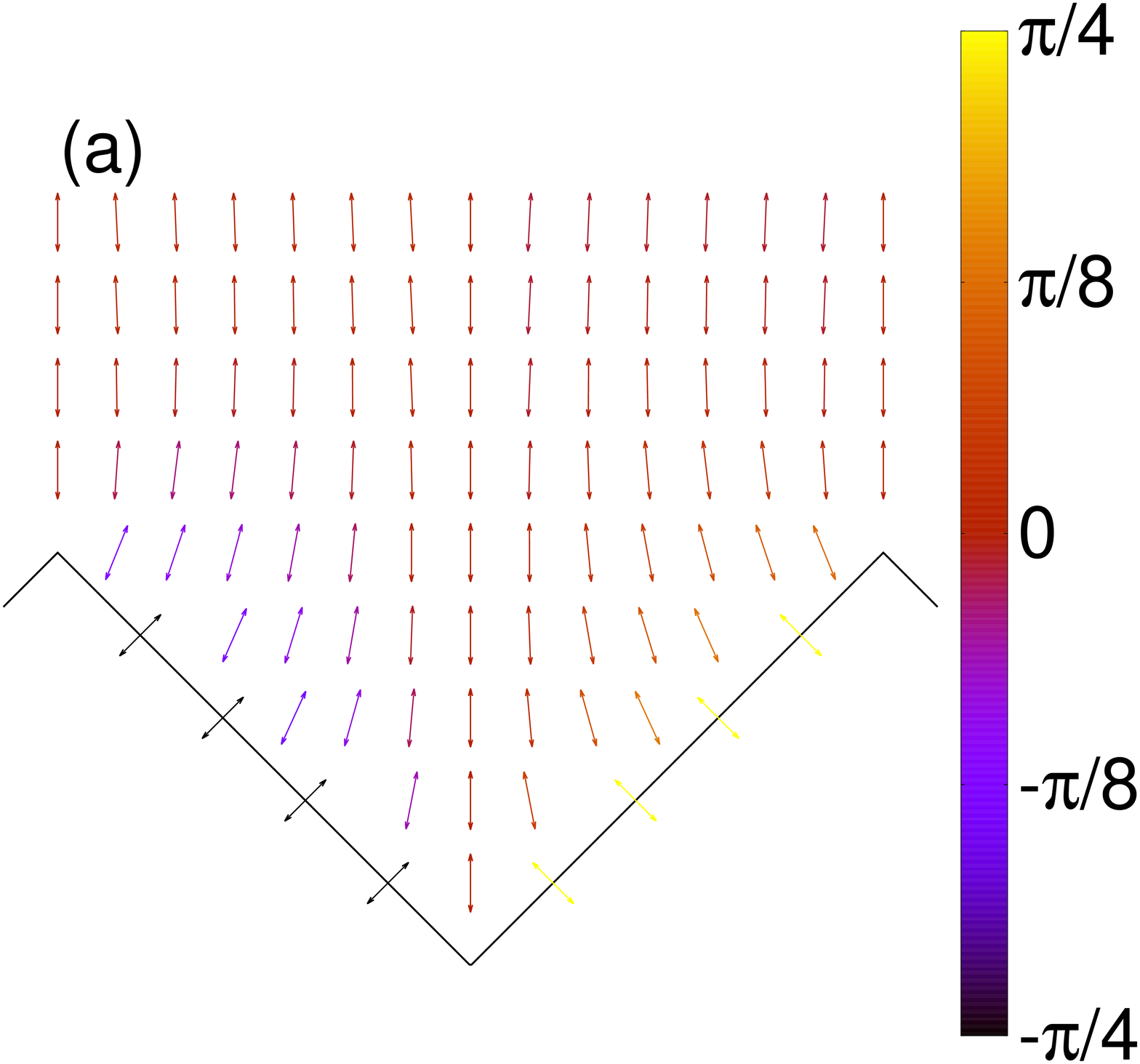} & 
\includegraphics[width=0.45\columnwidth]{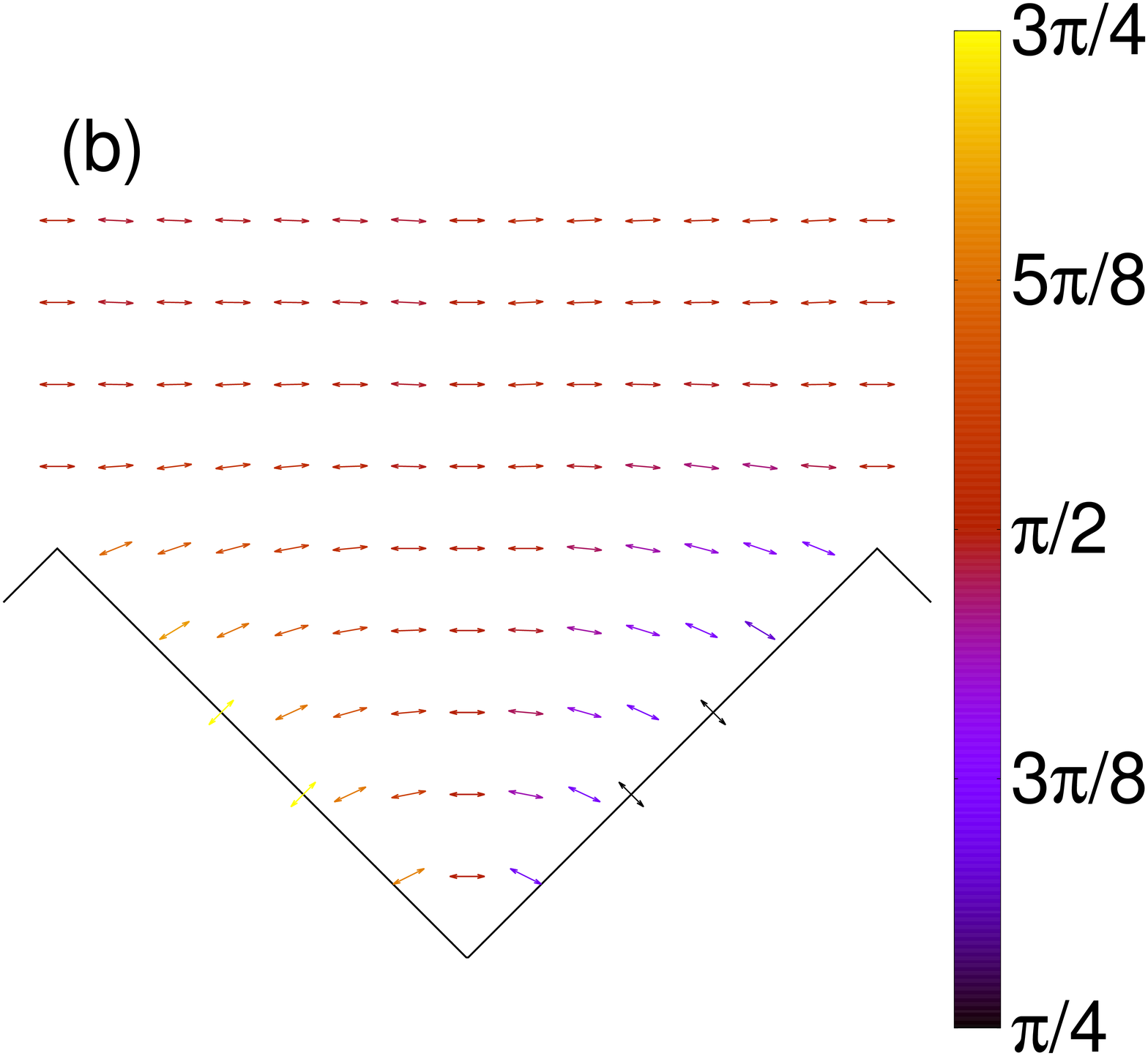}\\
\end{tabular}
\caption
{Coexisting textures at a sawtooth substrate ($\alpha=\pi/4$): 
(a) $N^\perp$ texture, (b) $N^\parallel$ texture. 
The arrows denote the local nematic director orientation and the colour code gives the 
orientational field $\theta$.
\label{fig5}}
\end{figure}
\clearpage
\vspace*{0.5cm}
\begin{figure}[H]
\centerline{
\includegraphics[width=0.8\columnwidth]{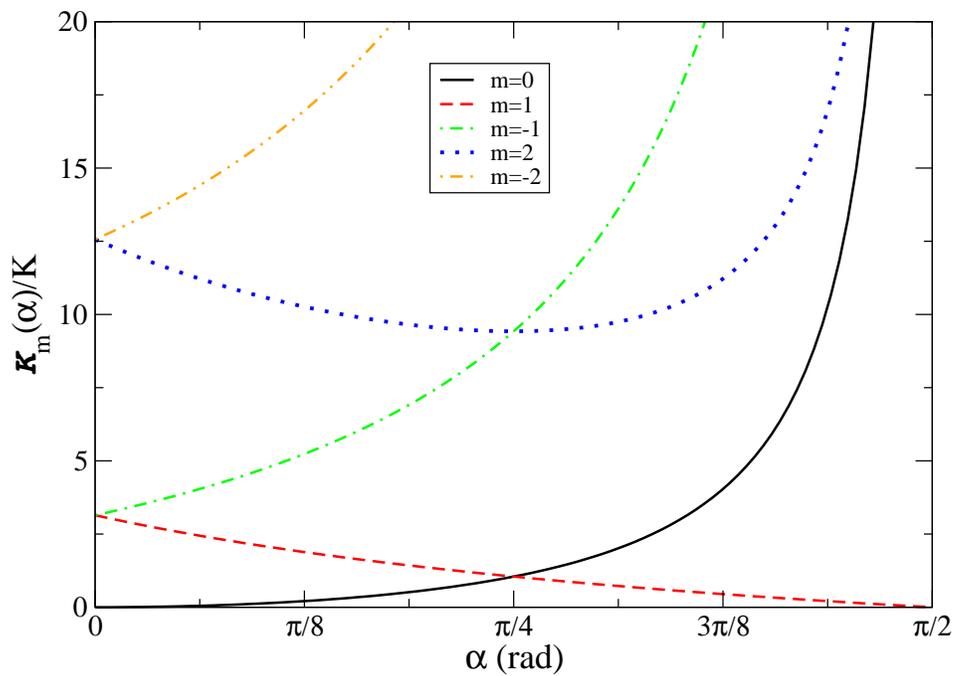}
}
\caption
{Plot of $\mathcal{K}_m/K$ as a function of $\alpha$ for different values of $m$: $m=0$ (continuous line),
$m=+1$ (dashed line), $m=-1$ (dot-dashed line), $m=+2$ (dotted line) and $m=-2$ (double dot-dashed line).
}
\label{fig6}
\end{figure}
\clearpage
\vspace*{0.5cm}
\begin{figure}[H]
\begin{tabular}{cc}
\includegraphics[width=0.45\columnwidth]{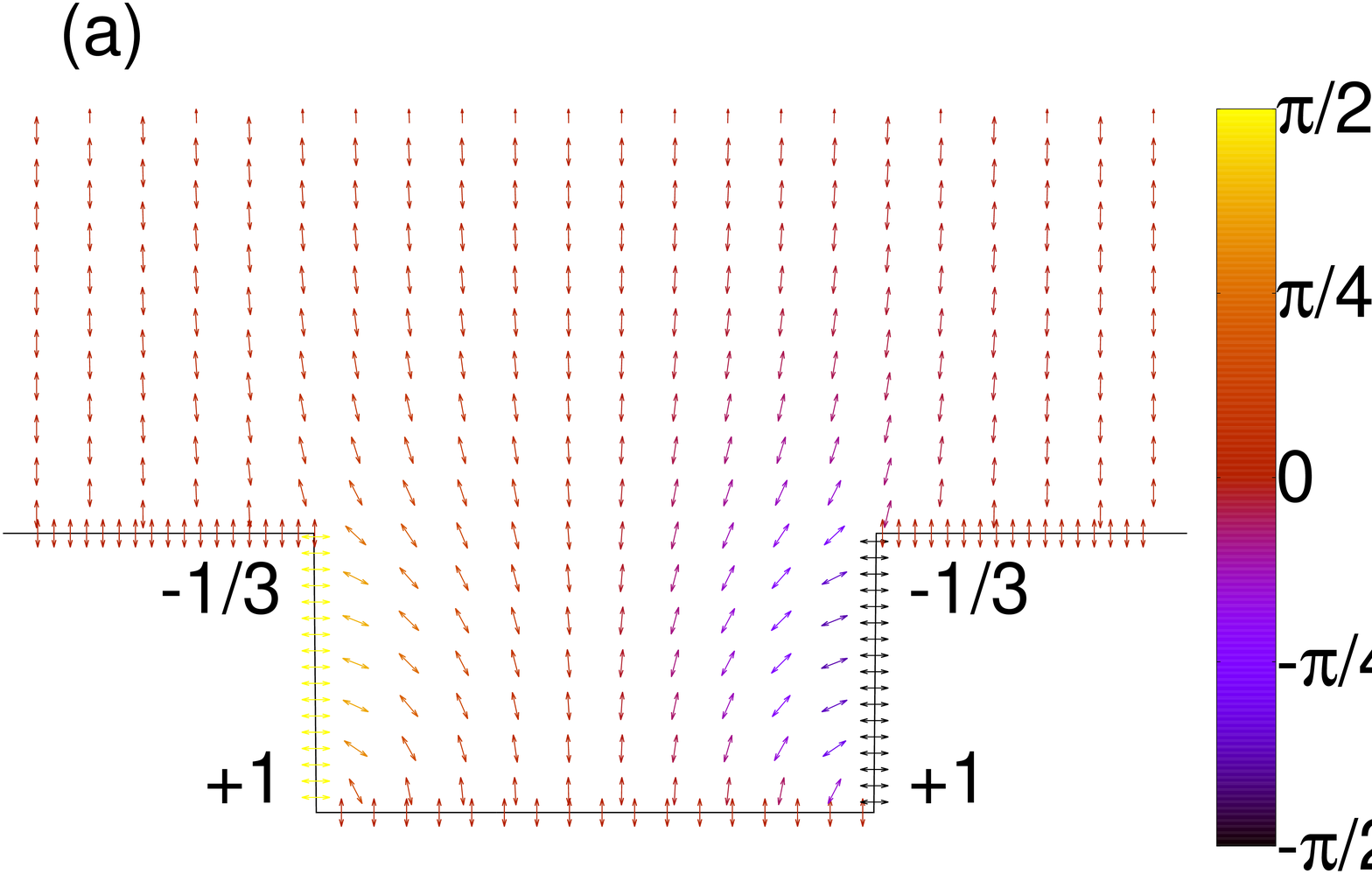} & 
\includegraphics[width=0.45\columnwidth]{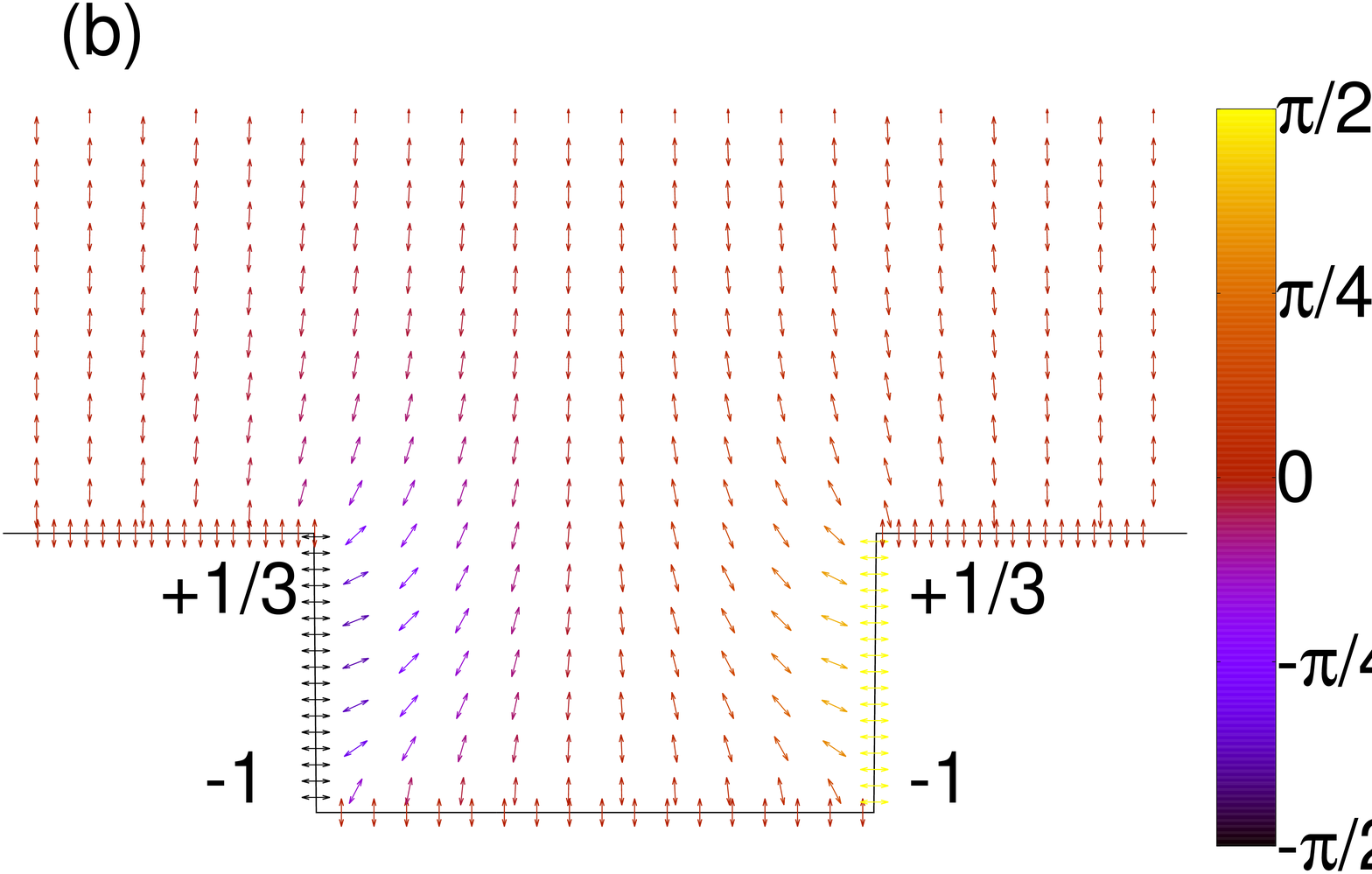}\\
\includegraphics[width=0.45\columnwidth]{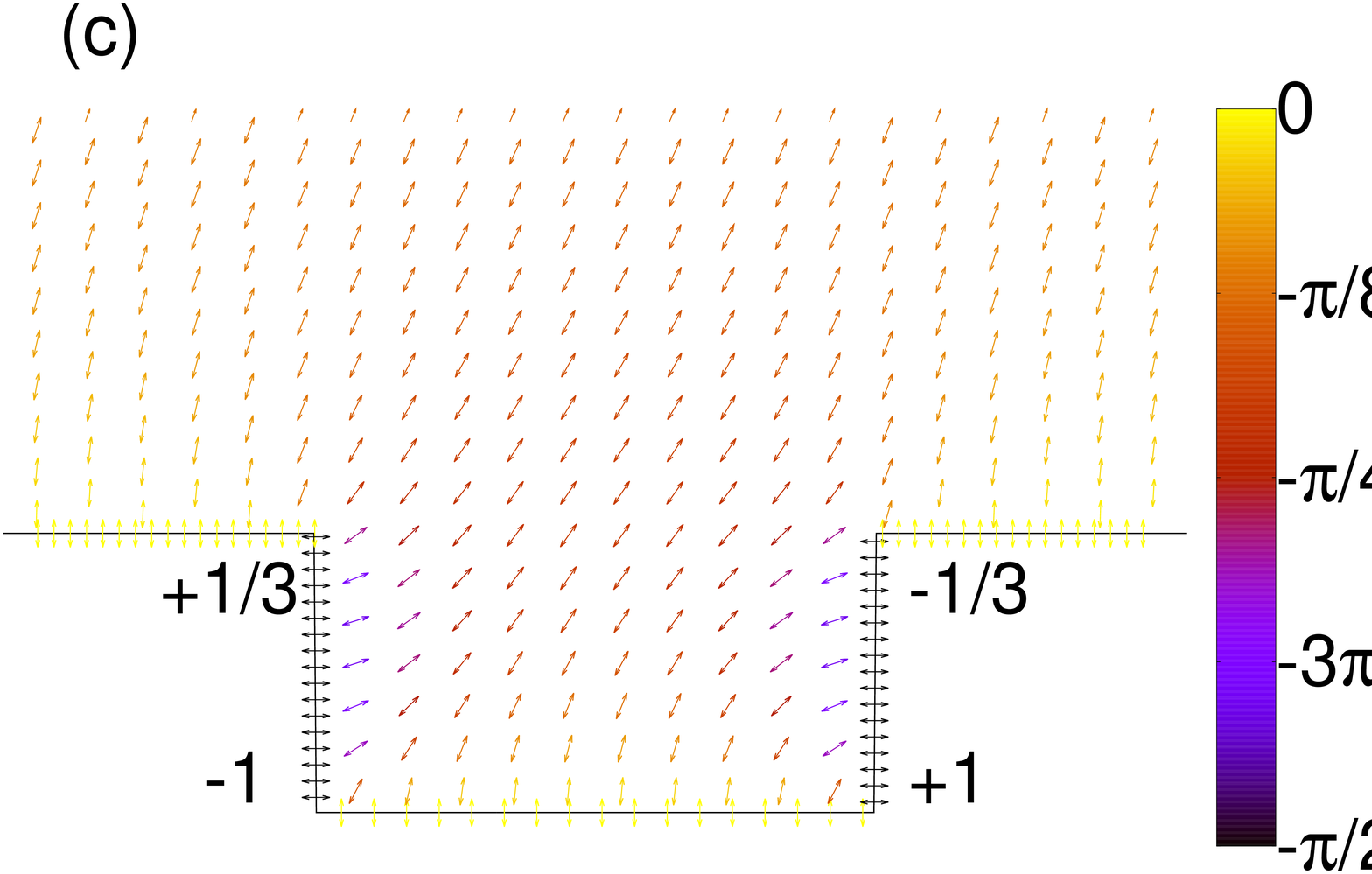} & 
\includegraphics[width=0.45\columnwidth]{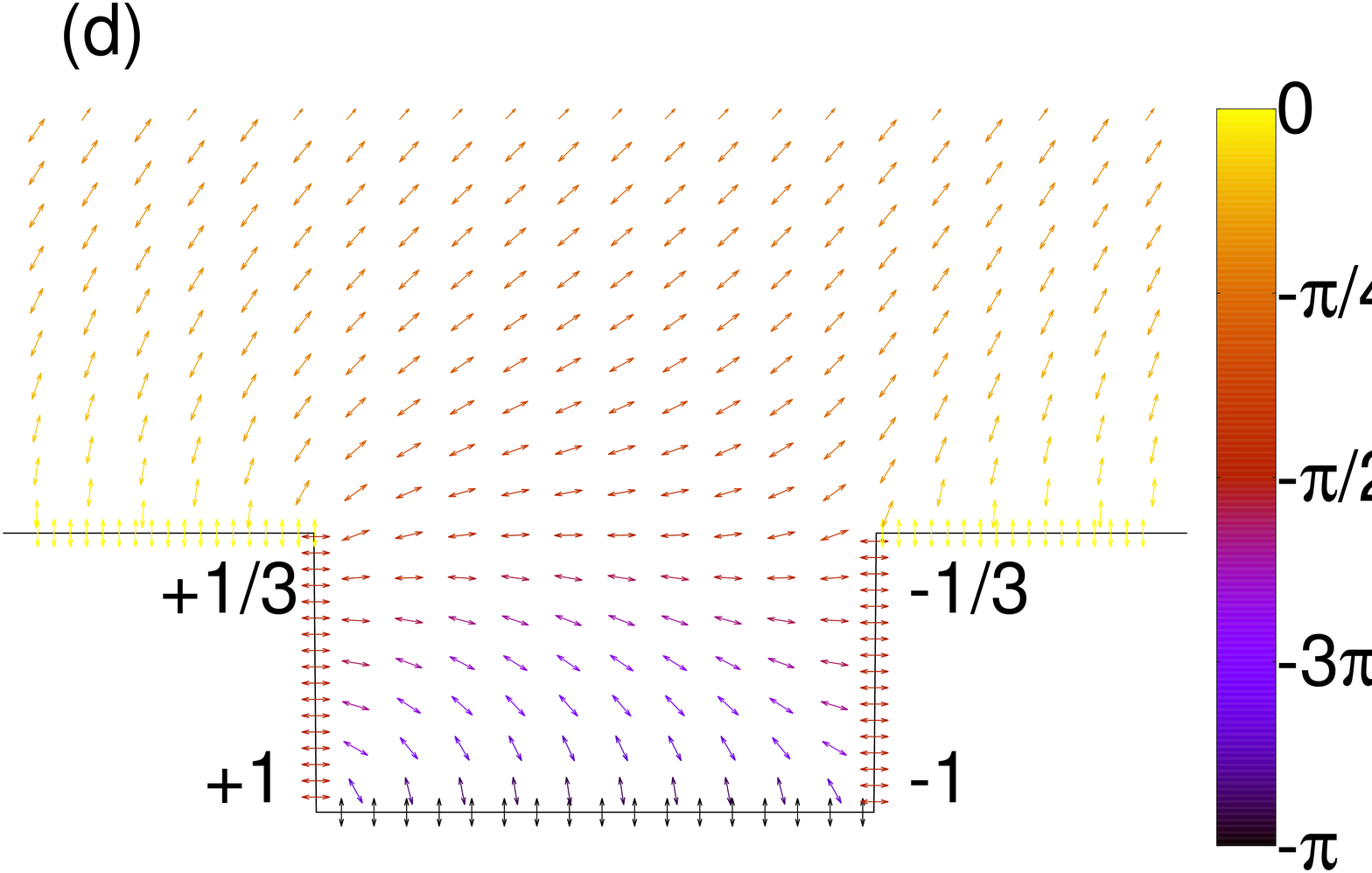}\\
\end{tabular}
\caption
{Typical textures at a crenellated substrate ($h/l_2=0.5$, $l_1/l_2=1$): (a) $N^\perp_1$ texture, (b) $N^\perp_2$ texture, 
(c) $N^o_1$ texture
and $N^o_2$ texture. 
The meaning of the symbols is the same as in Fig. \ref{fig5}.
\label{fig7}}
\end{figure}
\clearpage
\vspace*{0.5cm}
\begin{figure}[H]
\centerline{\includegraphics[width=0.8\columnwidth]{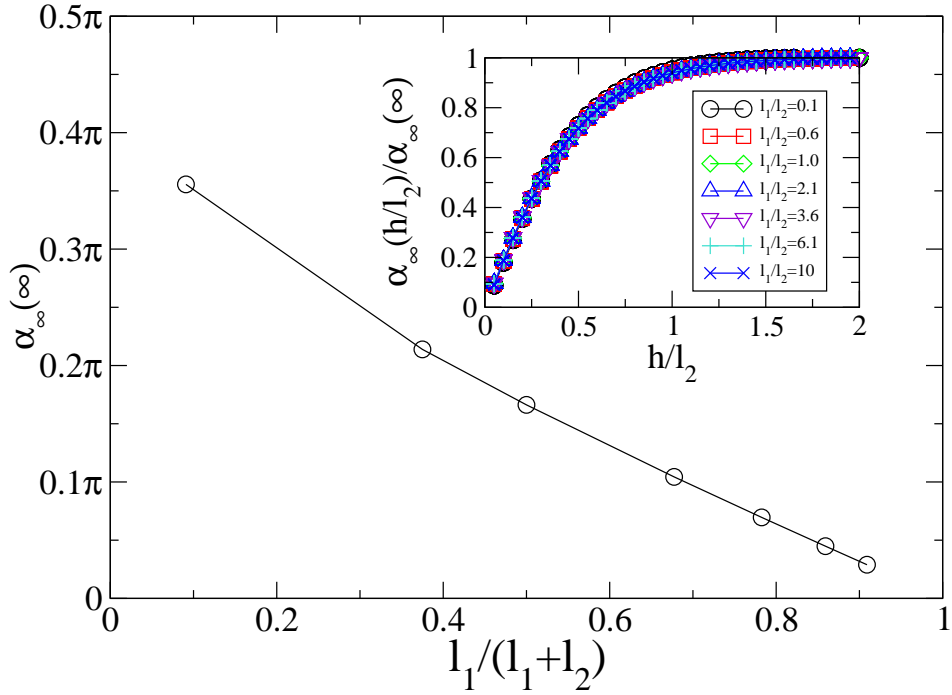}}
\caption
{Plot of the asymptotic far-field orientation $\alpha_\infty(h/l_2\to \infty$ as a function of $l_1/l_2$.
Inset: plot of the ratio $\alpha_\infty(h/l_2;l_1/l_2)/\alpha_\infty(\infty;l_1/l_2)$ as a function of $h/l_2$
for $l_1/l_2=0.1$ (circles), $0.6$ (squares), $1.0$ (diamonds), $2.1$ (triangles up), $3.6$ (triangles down),
$6.1$ (pluses) and $10$ (crosses).}
\label{fig8}
\end{figure}
\clearpage
\vspace*{0.5cm}
\begin{figure}[H]
\centerline{
\includegraphics[width=0.8\columnwidth]{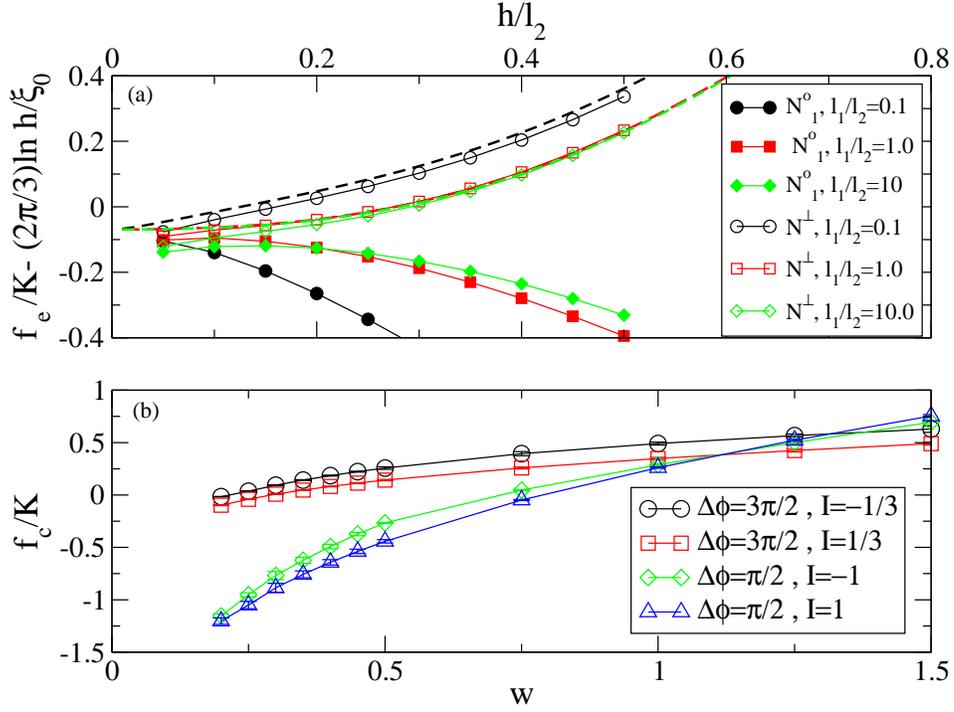}
}
\caption
{Top panel: Plot of the next-to-leading elastic contribution to the interfacial free energy with respect to 
$h/l_2$ corresponding to the $N^\perp_i$ (open symbols) and $N^o_1$ (filled symbols) for $l_1/l_2=0.1$
(circles), $1$ (squares) and $10$ (diamonds). Dashed lines correspond to the analytical prediction 
Eq. (\ref{mfofreeenergy2}). Bottom panel: plot of the core contributions as a function of the anchoring
parameter $w$ for the cusps with opening angle $\Delta \phi$ and topological
charge: $\Delta \phi=3\pi/2$ and $I=-1/3$ (circles), $\Delta \phi=3\pi/2$ and
$I=+1/3$ (squares), $\Delta \phi=\pi/2$ and $I=-1$ (diamonds) and 
$\Delta \phi=\pi/2$ and $I=1$ (triangles).  
}
\label{fig9}
\end{figure}
\clearpage
\vspace*{0.5cm}
\begin{figure}[H]
\centerline{\includegraphics[width=0.8\columnwidth]{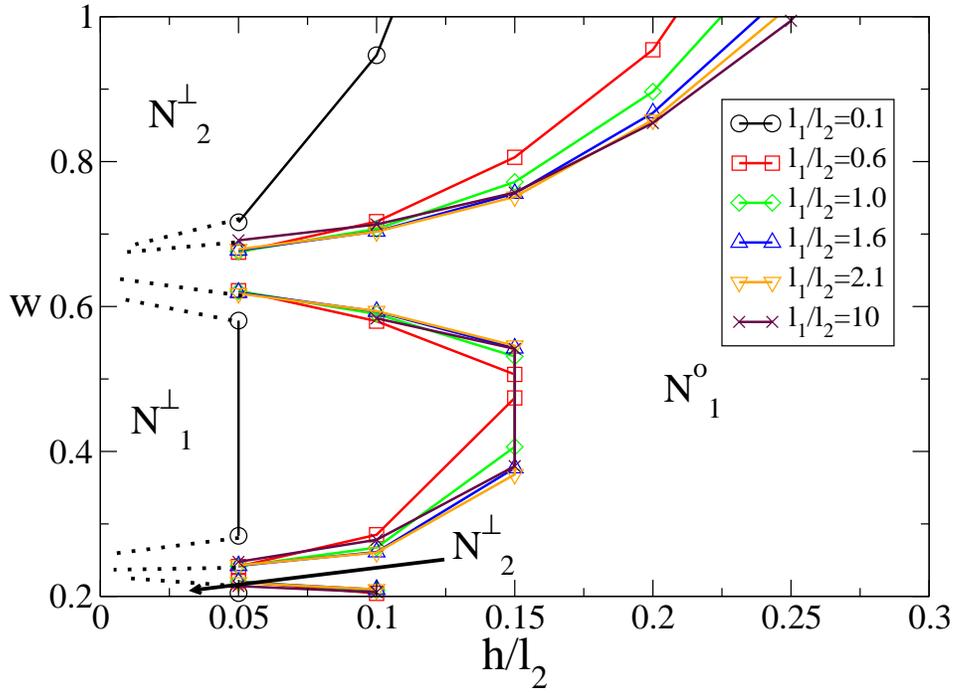}}
\caption
{Phase diagram for the crenellated substrate in terms of $h/l_2$
and the anchoring parameter $w$ for $l_1/l_2=0.1$ (circles), $0.6$ (squares),
$1.0$ (diamonds), $1.6$ (up-triangles), $2.1$ (down triangles) and $10.0$ (crosses). 
The lines are a guide to the eye, with the dotted lines illustrating the continuation
of coexistence at small $h$.}
\label{fig10}
\end{figure}
\clearpage
\vspace*{0.5cm}
\begin{figure}[H]
\begin{tabular}{cc}
\includegraphics[width=0.45\columnwidth]{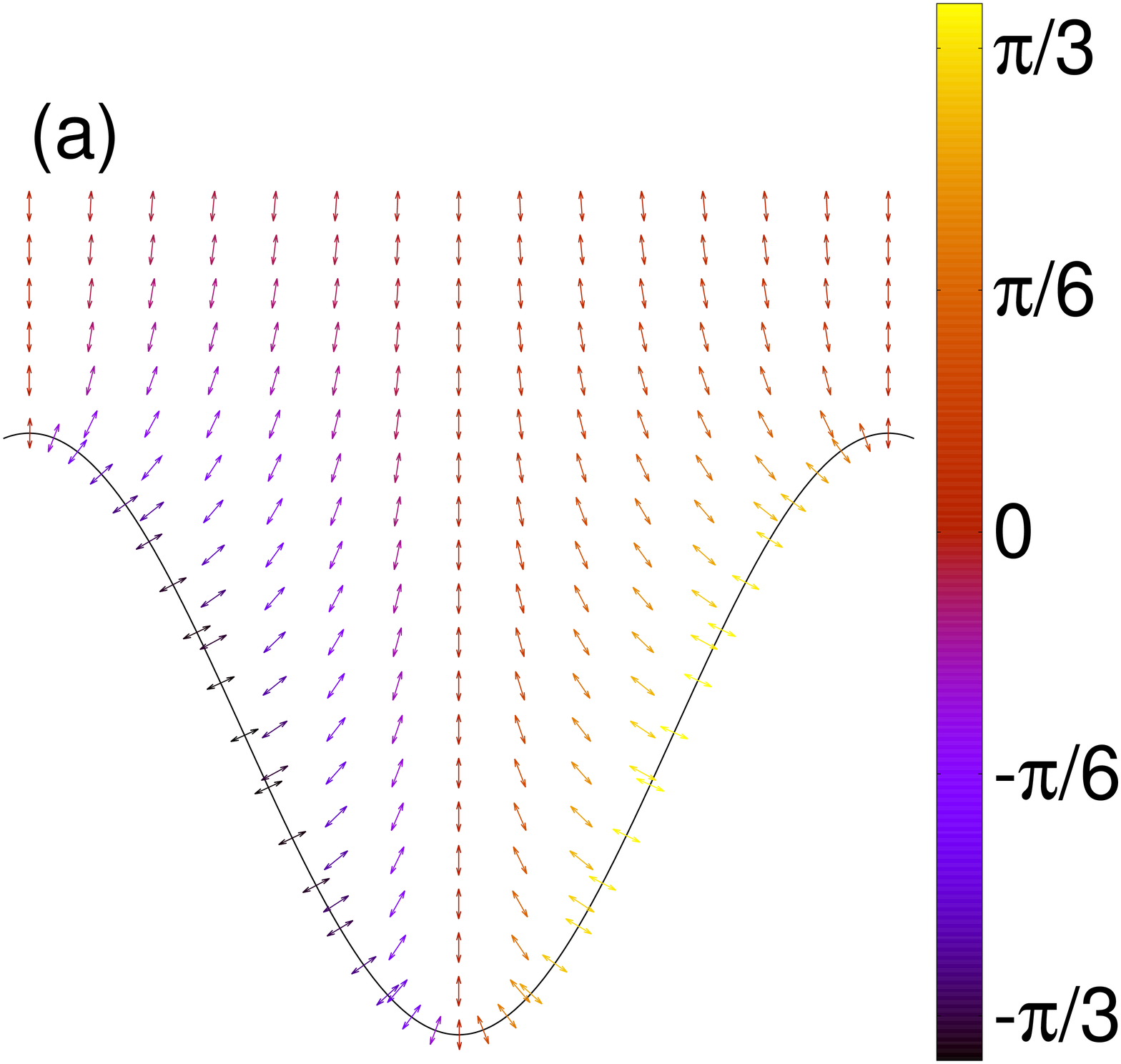} & 
\includegraphics[width=0.45\columnwidth]{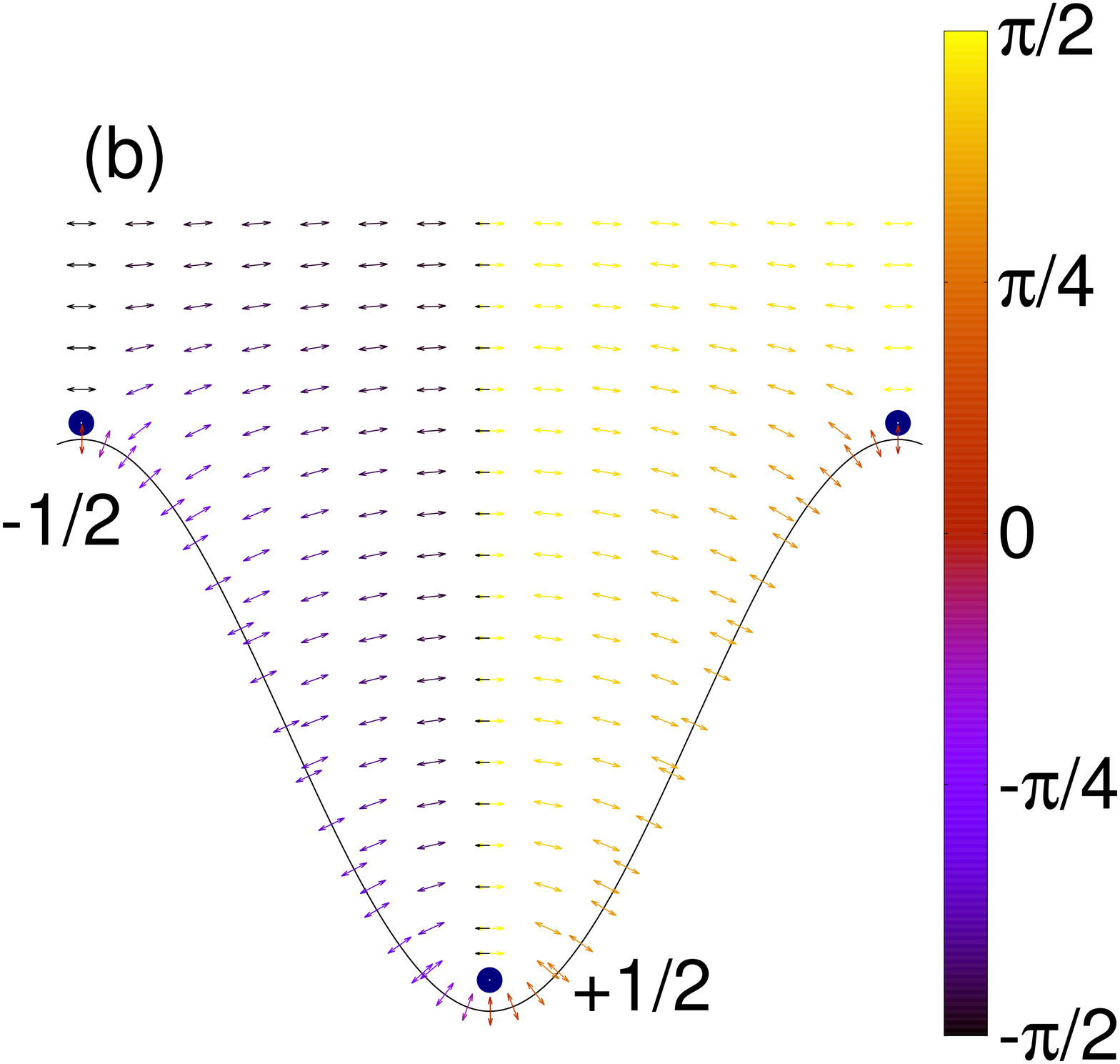}\\
\end{tabular}
\caption
{Typical textures at a sinusoidal substrate ($A/\lambda=0.35$): 
(a) $N^\perp$ texture, (b) $N^\parallel$ texture. In the latter, the position
of the disclination lines in the nematic are highlighted. 
The meaning of the symbols is the same as in Fig. \ref{fig5}.
\label{fig11}}
\end{figure}
\clearpage
\vspace*{0.5cm}
\begin{figure}[H]
\centerline{\includegraphics[width=0.8\columnwidth]{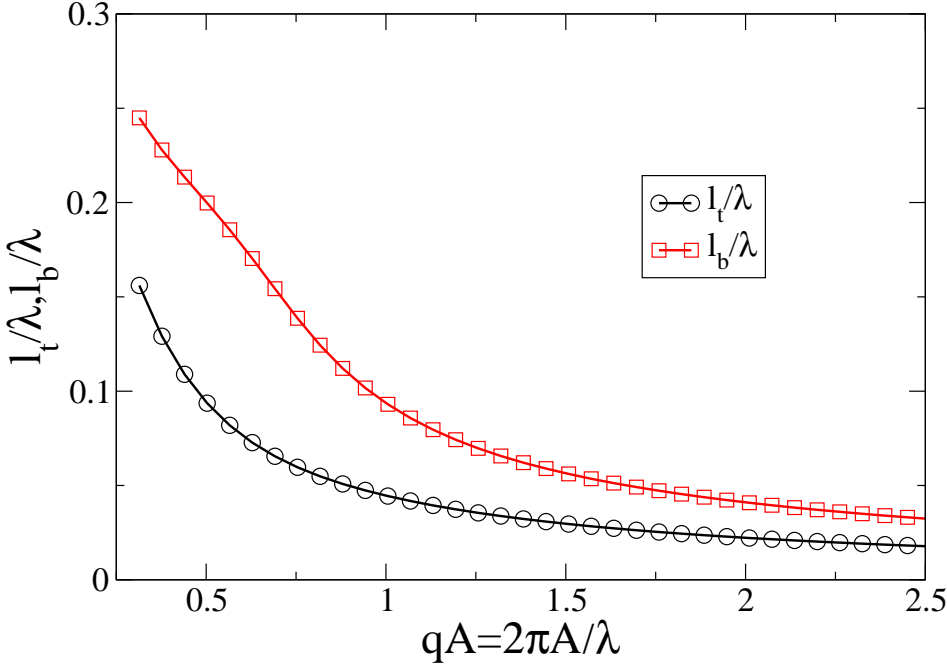}}
\caption
{Distance with respect to the substrate of the top and bottom disclination 
lines, $l_t$ and $l_b$, respectively, as a function of $qA$.}
\label{fig12}
\end{figure}
\clearpage
\begin{figure}[H]
\vspace*{0.5cm}
\centerline{\includegraphics[width=0.8\columnwidth]{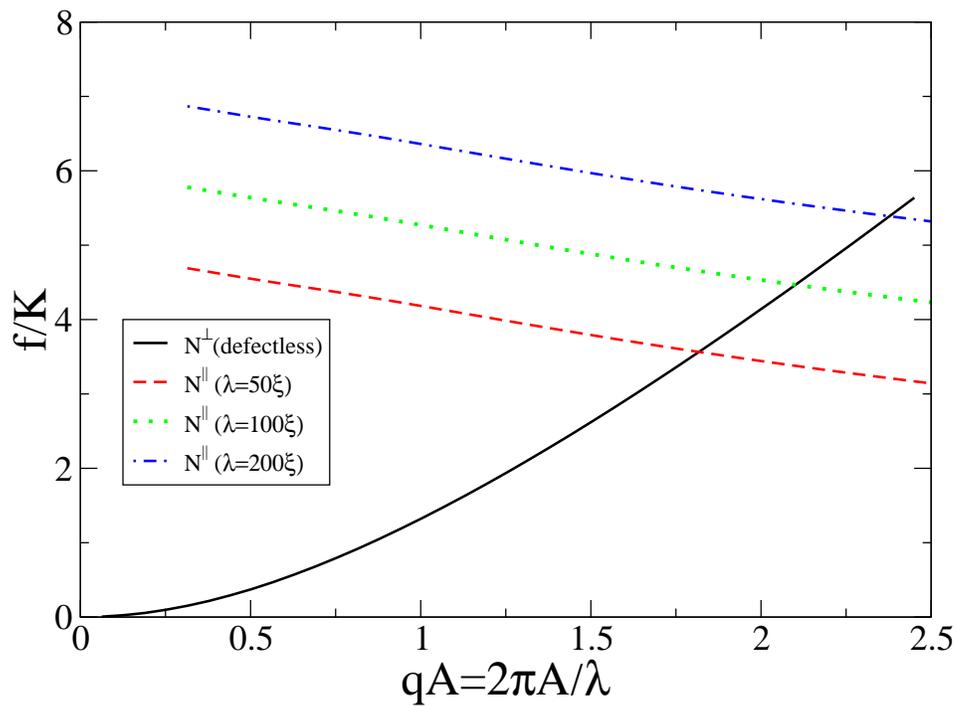}}
\caption
{Free energy of the $N^\perp$ (continuous line) and 
$N^\parallel$ textures with $\lambda/\xi_0=50$ (dashed line), $100$ (dotted
line) and $200$ (dot-dashed line). 
}
\label{fig13}
\end{figure}
\clearpage
\vspace*{0.5cm}
\begin{figure}[H]
\centerline{\includegraphics[width=0.8\columnwidth]{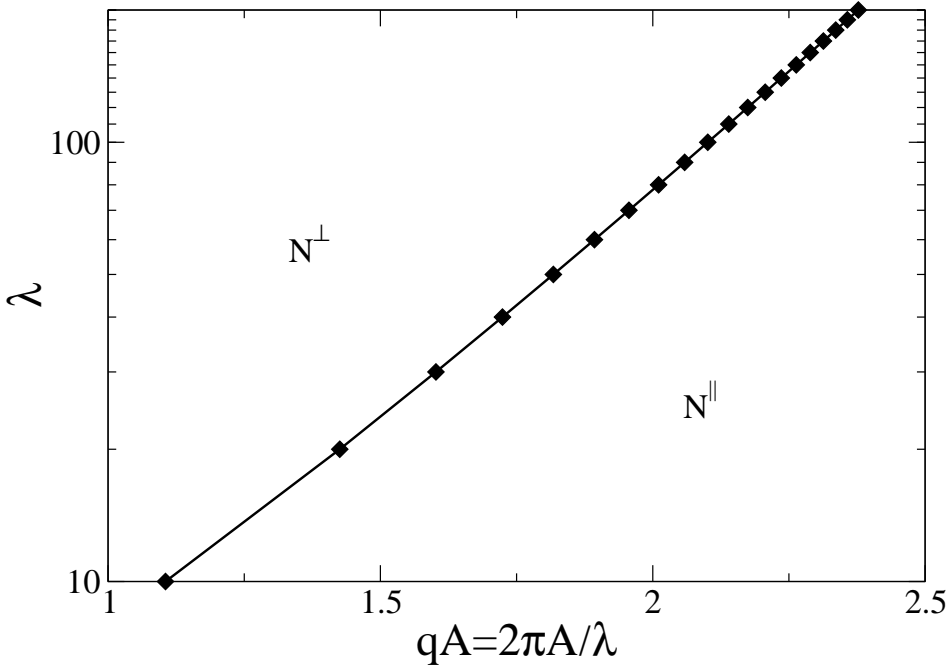}}
\caption
{Phase diagram for the crenellated substrate in terms of the roughness 
parameter $qA$ and the substrate period $\lambda$.}
\label{fig14}
\end{figure}
\clearpage
\vspace*{0.5cm}
\begin{figure}[H]
\centerline{\includegraphics[width=\columnwidth]{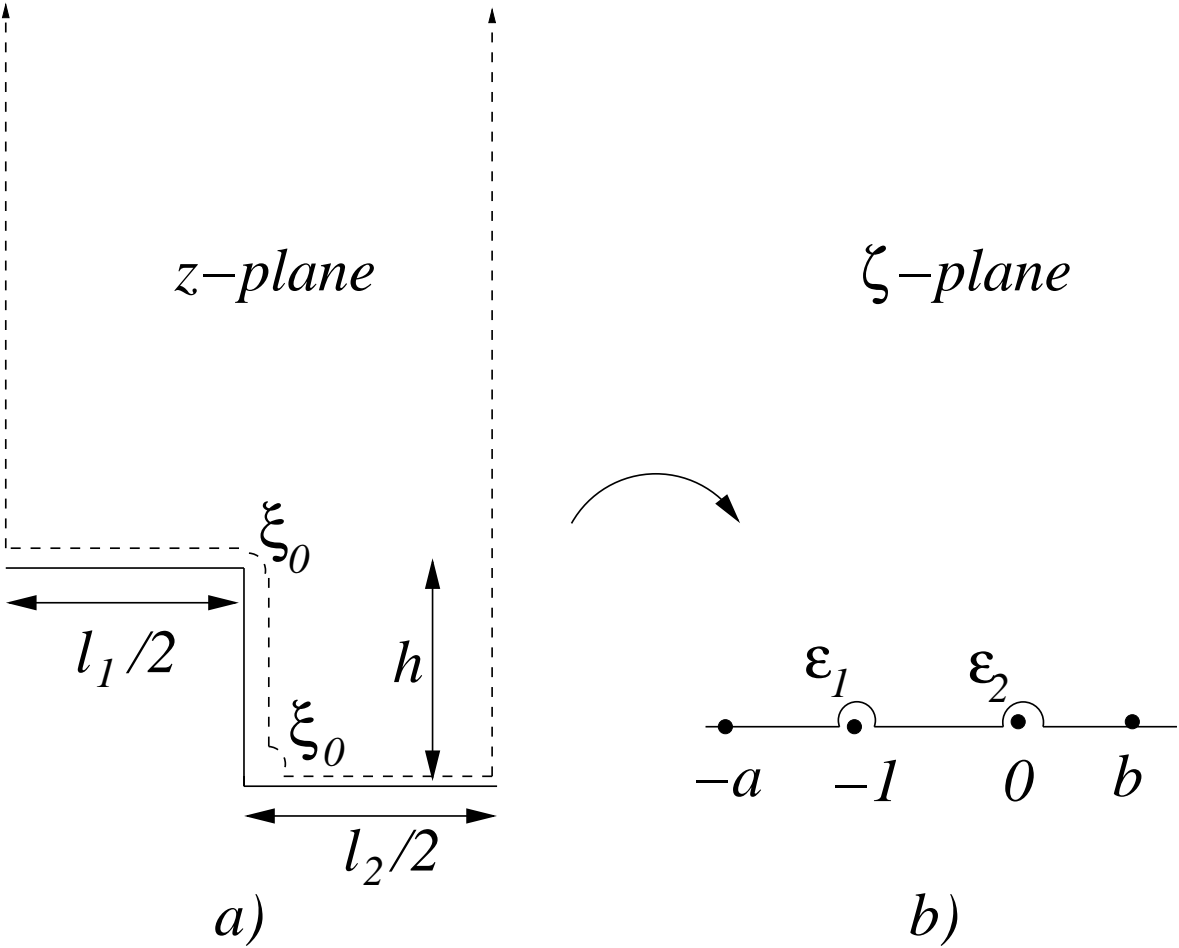}}
\caption
{Left panel: Minimization cell for the evaluation of the elastic free energy. The solid line
corresponds to the substrate, and the dashed line to the boundary of the domain where the orientation
field $\theta$ is calculated. Right panel:
Mapping of the minimization cell under the Schwarz-Christoffel transformation.}
\label{fig_appendix}
\end{figure}

\end{document}